\documentclass[12pt]{article}
\usepackage[utf8]{inputenc}
\usepackage{amsmath}
\usepackage [a4paper,top = 2cm, bottom = 2cm, left = 2.5cm, right = 2.5cm]
{geometry}
\usepackage{setspace}
\usepackage{graphicx}
\usepackage{comment}
\usepackage[]{subfigure}
\usepackage[font=scriptsize]{caption}
\usepackage[authoryear]{natbib}
\usepackage{xcolor}

\title{Better to stay apart: asset commonality, bipartite network centrality, and investment strategies}

\begin{document}

\author{
{Andrea Flori\thanks{Politecnico di Milano, Department of Management, Economics and Industrial Engineering; andrea.flori@polimi.it}}
\and {Fabrizio Lillo\thanks{Center for Analysis, Decisions, and Society (CADS) - Human Technopole, Milano, Italy; fabrizio.lillo@unibo.it} \thanks{Università di Bologna, Department of Mathematics, Bologna, Italy}}
\and {Fabio Pammolli\thanks{Politecnico di Milano, Department of Management, Economics and Industrial Engineering; fabio.pammolli@polimi.it} \thanks{Center for Analysis, Decisions, and Society (CADS) - Human Technopole, Milano, Italy}}
\and {Alessandro Spelta\thanks{Center for Analysis, Decisions, and Society (CADS) - Human Technopole, Milano, Italy; alessandro.spelta@htechnopole.it}}}

\date{}
\maketitle

\begin{abstract}
By exploiting a bipartite network representation of the relationships between mutual funds and portfolio holdings, we propose an indicator that we derive from the analysis of the network, labelled the \textit{Average Commonality Coefficient} ($ACC$), which measures how frequently the assets in the fund portfolio are present in the portfolios of the {\it other} funds of the market.  This indicator reflects the investment behavior of funds' managers as a function of the popularity of the assets they held. We show that $ACC$ provides useful information to discriminate between funds investing in niche markets and those investing in more popular assets. More importantly, we find that $ACC$ is able to provide indication on the performance of the funds. In particular, we find that funds investing in less popular assets generally outperform those investing in more popular financial instruments, even when correcting for standard factors. Moreover, funds with a low $ACC$ have been less affected by the 2007-08 global financial crisis, likely because less exposed to fire sales spillovers.


\end{abstract}

\newpage

\section{Introduction}
\label{intro}
\doublespacing

Evaluating funds' performances is of major interest for investors and market efficiency in general. Scholars have proposed several alternative models to both explain funds' performances and identify the main factors driving extra-performances. Starting from the traditional asset pricing model (namely, \textit{CAPM}) which evaluates stocks performances in terms of how they are related to market returns, literature has introduced additional factors aimed at identifying peculiar risk contributors. \cite{fama1993common} discussed a linear three-factors model where, in addition to market premium, two further factors were discussed to measure the historical excess returns of small vs. big caps and value vs. growth assets. \cite{carhart1997persistence} enriched this framework by proposing a four-factors model where a momentum factor was defined to capture the role of \textit{winner minus loser} assets in the market. More recently, \cite{fama2015five} extended their three-factors model by adding profitability and investments factors, while \cite{pastor2002mutual} proposed a seven-factors model where three industry factors were added into the Carhart's model.

Market players are, however, likely to interpret information in different manners and actively manage their portfolios in the pursuit of generating performances which beat those expected from these factors. Skilled managers are those investors that should be more able to extract market signals and invest accordingly, thus generating positive extra-performances than less skilled investors. This, in turn, calls for the identification of systematic patterns in the way investors produce these extra-performances (namely, \textit{alphas} in financial jargon), and whether they are persistent in time. Literature discussed to what extent these investors, that are considered as skilled players, are more likely to generate positive extra-performances and, eventually, beat other investors systematically. However, the predictability of funds performances has been questioned by empirical evidence and this fact motivated the growing literature on the relationship between managerial skills and the persistence of alphas (see e.g., \cite{grinblatt1992persistence,hendricks1993hot,goetzmann1994winners,brown1995performance,elton1996persistence,bollen2004short,barras2010false,busse2010performance,fama2010luck}, among others).

One important determinant of alpha is clearly the ability to choose potentially profitable assets. It is likely that assets that are chosen by many other funds contribute to profitability in a way that is already explained by existing factors. For this reason in this paper we investigate whether the ability of picking assets which are not common to other funds brings information on  managerial skills that is not embedded in traditional alpha measures.

In order to perform this analysis, we look at the system of funds and assets from a global point of view. The system can be seen as a bipartite network, where a link between a node in one set (funds) and a node in the other (assets) indicates a significant investment of the fund in the considered asset (see subsection \ref{method} for a more formal definition). The similarity between two portfolios, termed the overlap\footnote{Portfolio overlaps are receiving increasing attention, especially in the literature of systemic risk due to fire sales spillover. See, for example, \cite{caccioli2014stability, Greenwood,Corsi,DiGangi}.}, is broadly defined as the fraction of common assets, while the commonality of an asset is the number of funds owning it. We can therefore define the {\it Average Commonality Coefficient} ($ACC$) of a fund as the average commonality of the assets in its portfolio.
Here we are thus interested in identifying not only those funds that diversify the most in terms of portfolio composition, but also we can recognize either those assets that are present in a huge share of funds or, alternatively, those that are held by few portfolios only. Hence, given the same level of diversification, as naively measured for instance by the number of assets in the portfolio, we can discriminate between those funds more prone to invest in niche markets (namely, in assets not present in many portfolios) and those that opt for common assets (namely, assets quite well-spread and \textit{popular} among funds). 

In particular, in this paper we focus on US equity mutual funds and we investigate managerial skills by focusing on the topological features of portfolio holdings\footnote{In the paper the terms constituents and assets, or holdings and compositions are considered interchangeable.}. Compositions may be, in fact, informative in signaling managerial preferences and changes in portfolio holdings can be exploited to reveal future funds' performances and managers' attitudes to risk. For instance, during market uncertainty, such as the crash of mid-2007, investors allocated a growing portion of their portfolios to safer assets (see e.g., \cite{kacperczyk2010safe,rosch2013market,bethke2017investor} among others), corroborating a \textit{flight to quality} selection of assets at portfolio level that might have impacted on the overall similarity across funds and possibly to the effective extent of diversification in the market.

This work relates the extra-performance of each fund to the level of similarity with the rest of the system, where the latter combines both how the fund manager diversifies the portfolio and how the assets he or she selects are also selected by many other fund managers. In this regards, we do not rely on a typical indicator of diversification, e.g. the level of concentration of assets under management in the fund, but we apply a novel measure (the $ACC$ indicator) which embeds also the popularity of the assets present in the portfolio. This measure is related to a well-diffused toolbox of indicators developed in economic complexity by \cite{hidalgo2007product,hidalgo2009building} to assess the topological structure of an economic system, such as the degree of diversification and the extent of specialization of each agent operating in that system. The economic-complexity index (ECI) developed by Hidalgo and Hausmann has been exploited to predict the future economic growth of a country by looking at diversity and sophistication of the products such country exports. Essential to the notion of economic complexity are analyses of the interconnected patterns of countries’ trade exports in the global ``product space''. In this article, we start from these concepts and we investigate how funds move in the ``portfolio constituents space''. The essential insight is that constituents selected by a fund represent a proxy measure for the managerial ``capabilities'' of the manager. 

This approach is, in part, in line with \cite{cohen2005judging}, who already recognized that similarity among portfolio holdings can provide useful information for performance predictability that is usually not included in alpha measures. In particular, their approach maps similarity across managers by measuring the \textit{quality} of the assets held in their portfolio according to a weighting scheme that is based on the average alpha of the managers that invest in these assets. Here, by contrast, we attempt to gauge a different perspective of portfolio diversification which basically takes into account also the choice made by managers to pick niche vs. popular assets in the pursuit of positive alpha. In the following we show, that $ACC$ is weakly correlated with the one of \cite{cohen2005judging}, and that gives different and often better predictions of fund's performance.

We do not discuss why investors opt for more or less niche portfolios or, in our perspective, for more or less peripheral assets. We present, instead, empirical evidence that managerial ability to generate extra-performances reflects the commonality properties of the assets under management and that this effect is in part affected by the impact of the crisis of mid-2007. Generally speaking, we observe that the $ACC$ dimension does not emerge simply as a proxy for managerial skills, but rather as a complementary criterion to alpha measures for building profitable investment strategies.

Literature already presented empirical evidence that the performances of actively managed funds relate to the way they concentrate their portfolios according to their informational advantages (see e.g., \cite{coval1999home,kacperczyk2005industry,cremers2009active}). Here we  exploit the bipartite network topological structure to evaluate whether managers' extra-performances can be related to a different investment attitude towards asset commonality. This proposed indicator of manager's skill is exploited to describe alpha persistence in time and to interpret funds' extra-performances during the market turmoil of the recent global financial crisis. We find that, after controlling for three and five factors, those funds with more peripheral assets (namely, those funds with low values for $ACC$) are more prone to produce positive extra-performances than those investing in more popular assets. Portfolio strategies investing long in funds with low values of $ACC$ and short in those funds with high values for this topological indicator are then able to generate positive extra-performances even along an holding period characterized by a boom and burst cycle.

The rest of the paper is organized as follows: Section \ref{datam} will present the data set and the methodology applied to compute the $ACC$ indicator, discussing in particular how the network theory representation of the mutual funds perimeter can be exploited to extract information from portfolio holdings; Section \ref{results} shows the results of our investment strategies involving the topological information under different time windows across the crisis; then, Section \ref{discussion} discusses the main economic implications from the use of $ACC$ against alternative formulations of measures for managerial skills. Finally, Section \ref{conclusion} concludes.

\section{Data and Methodology}
\label{datam}
\subsection{Data}
\label{data}

Data are retrieved from the CRSP Survivor Bias-Free data set which collects historical holdings and performances for US open-ended mutual funds. CRSP database provides a mapping between portfolios and the funds investing on them. For instance there could be the case that a certain portfolio is held by multiple funds. In order to study portfolio overlapping across funds, we assign funds' gross returns\footnote{Following \cite{cohen2005judging} we include the annual expense ratio and 12(b)1 fees given by CRSP; we divide these amounts by 252 to get daily quota, and we add the resulting value to each daily net fund's return to obtain gross returns.} to the corresponding portfolios proportionally to the funds total net asset values. By doing this, hereinafter terms funds and portfolios are used interchangeably. To study the relationships between funds and their constituents, we focus on those funds more involved in US equity instruments. This selection has been performed, in line with \cite{schwarzkopf2010}, by taking those funds with equity exposures corresponding to at least 80 per cent of the net asset value of the portfolio. 
Our data set encompasses portfolio holdings from 2004 to 2010 at a quarterly basis, while constituents' and funds' returns are mapped daily. The analyses are also conducted on two sub-samples, the pre-crisis period that ranges from 2004 to 2007 and the (post)-crisis period, from 2007 to 2010.

Starting from the raw data we have aggregated constituents using the Cusip ticker and we have added funds' fees so to obtain gross funds' returns. The overall data set includes more than 2,700 funds investing in about 15,000 constituents, whose averages are 1,882 and 10,274, respectively. Both the number of funds and of constituents present in the data set increase along time, from the year 2004 where we observe 1,113 funds and 5,018 constituents to the year 2010 where 2,345 funds and 14,334 constituents are collected in the sample. Despite the large number of entities recorded in the data set, not all funds and constituents are persistently present at all the releasing dates. At each quarter we consider in our analysis 
only to those funds (and corresponding constituents) that are present consecutively in two quarters. 

Finally, our study relies on the alpha measures of managerial skills obtained from the three-factors and five-factors models (\cite{fama1993common,fama2015five}), using time series retrieved from the K. French data library (http://mba.tuck.dartmouth.edu/pages/faculty/ken.french/data\_library.html).




\subsection{Methodology}
\label{method}

\begin{figure}[ht!]
\caption{\textbf{Example of the bipartite network for 100 funds and constituents.} The figure represents the relationships between the first 100 funds and the first 100 constituents (in alphabetical order) for the last quarter of 2005. Notice the heterogeneity between the links connecting funds and constituents. Some constituents are very popular (held by the majority of the funds) while others are present in few portfolios only. Similarly, some funds invest in many assets while other funds concentrate their portfolios in few assets.}
\centering
\includegraphics[scale=0.90]{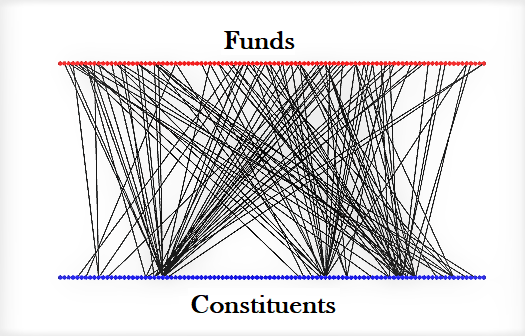}
\label{bipartito}
\end{figure}

Our data set can be easily interpreted as a dynamic bipartite network $H_{t}\left(F,C\right)$ in which nodes can be separated into two types, funds $(F)$ and constituents $(C)$, such that links only connect nodes in different partitions (for an example see Figure \ref{bipartito}). In recent years, many economic and financial systems have been described and modeled in terms of bipartite networks (see e.g., \cite{tumminello2011statistically,huang2013cascading,caccioli2014stability,barucca2016disentangling}).

In order to focus on constituents that really define the behavior of each fund, a stringent measure of portfolio composition is needed. We say that a fund $F_{i}$ holds a relevant exposure to constituent $C_{j}$ whenever the ratio of the market value of constituent $C_{j}$ in the portfolio of fund $F_{i}$ over the average market value of constituent $C_{j}$ in the whole galaxy of funds is greater than a certain threshold $x$. This definition is the analogous of the Revealed Comparative Advantage (RCA) proposed by \cite{balassa1965trade} and previously applied in the trade networks by \cite{hidalgo2007product,hidalgo2009building}.

For each quarter $t$ and fund $F_{i}$ we compute the relative holding $RH$ for constituent $C_{j}$ as follows:

\begin{equation*}
\centering
RH_{t}(F_{i},C_{j}) = \frac{H_{t}(F_{i},C_{j})}{\sum_{C_{j}}H_{t}(F_{i},C_{j})} \bigg/
\frac{\sum_{F_{i}}H_{t}(F_{i},C_{j})}{\sum_{F_{i}}\sum_{C_{j}}H_{t}(F_{i},C_{j})}
\end{equation*}

\begin{figure*}[!ht]
\centering
\caption{\textbf{Network Density for different threshold values.} Semi-Log plot of the percentage of links present for different threshold values. The dashed gray line identifies $RH_{t}\left( F_{i}.C_{j}\right)=1$ that represents the market average. Each line stands for one particular quarter. }
\includegraphics[width=12cm,height=7.5cm]{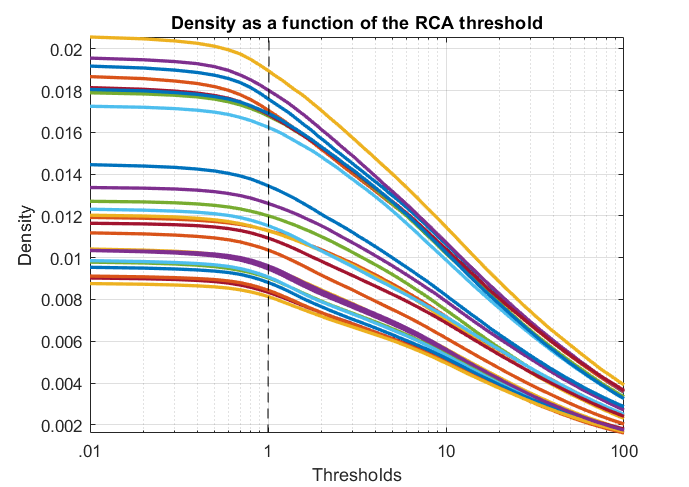}
	\label{fig:threshCusip}
\end{figure*}

This helps us in defining the bipartite network $M_{t}\left(F_{i},C_{j}\right) $ of funds' holdings at every quarter. We set $M_{t}\left( F_{i},C_{j}\right) =1$ if the relative holding of funds $F_{i}$ with respect to the constituent $C_{j}$ at time $t$ is greater or equal then $1$, i.e. $RH_{t}\left(F_{i},C_{j}\right) \geq 1$. This measure informs whether a fund's holding of a constituent is larger or smaller than the average holding of the entire galaxy of funds. We perform also some robustness analysis by letting the threshold to vary from 0 to 100. Figure \ref{fig:threshCusip} indeed shows the network density, i.e. the portion of the potential connections in a network that are actual connections as long as the threshold varies.

As in \cite{hidalgo2007product} and \cite{hidalgo2009building}, we now consider the temporal bipartite network $M_{t}$ described by the adjacency
matrix $M_{t}\left(F_{i},C_{j}\right) $, where $M_{t}\left(F_{i},C_{j}\right) =1$ if fund $F_{i}$ is connected to constituent $C_{j}$ and zero otherwise. Dropping the temporal suffix $t$, the method of reflections\footnote{There is a vivid debate on how to apply indicators of economic complexity in different economic fields (see e.g., \cite{bahar2014neighbors,gala2017economic,hartmann2017linking,desmarchelier2018product}). Furthermore, recent publications have also proposed nonlinear versions of the algorithm to measure centrality in bipartite networks; the interested reader can refer e.g. to: \cite{tacchella2012new,tacchella2013economic,morrison2017economic,alshamsi2018optimal}, among others.} consists of computing iteratively the average value of the previous-level properties of a node's neighbors and is defined as the set of observables:%
\begin{eqnarray*}
k_{F_{i},N} &=&\frac{1}{k_{F_{i},0}}\sum_{C_{j}}M\left(F_{i},C_{j}\right)
k_{C_{j},N-1} \\
k_{C_{j},N} &=&\frac{1}{k_{C_{j},0}}\sum_{F_{i}}M\left(F_{i},C_{j}\right)
k_{F_{i},N-1}
\end{eqnarray*}%
for $N\geq 1$, with initial conditions given by the degree, or number of links, of funds and holding constituents:%
\begin{eqnarray*}
k_{F_{i},0} &=&\sum_{C_{j}}M\left(F_{i},C_{j}\right) \\
k_{C_{j},0} &=&\sum_{F_{i}}M\left(F_{i},C_{j}\right)
\end{eqnarray*}

We can easily summarize the interpretation of the variables described by the method of reflections in economical terms. Indeed, ${\small k}_{F_{i},0}$ represents the number of constituents, i.e. the {\it diversification} of the fund holds\footnote{To be precise, the quantity ${\small k}_{F_{i},0}$ represents the number of constituents held by a fund whose holding is greater than or equal to the share held on average by the other funds.}. ${\small k}_{C_{j},0}$ is the number of funds having constituent $C_{j}$ in their portfolio (i.e., the commonality property). Basically, if ${\small k}_{F_{i},0}$ is low it means that fund $F_{i}$ is very concentrated in few assets, while a high ${\small k}_{F_{i},0}$ represents a fund that diversifies its portfolio among many assets; by contrast, a low ${\small k}_{C_{j},0}$ means that asset $C_{j}$ is a niche asset held by few funds whereas a high value of ${\small k}_{C_{j},0}$ represents a popular asset present in many portfolios. In the economic complexity jargon this feature refers to the commonality property of the node in the network, which basically in this context indicates how the asset is popular/common\footnote{Hence, the term commonality applied here is slightly different from the usage in other financial applications (see e.g., \cite{flannery1984effect,allen2012asset,namvar2013commonalities}, among others). } among portfolio holdings.

Recursively, the variable ${\small k}_{F_{i},1}$ is the average commonality of constituents in the portfolio of fund $i$, while ${\small k}_{C_{j},1}$ represents the average diversification of the funds having constituent $C_{j}$ in their portfolio. Since we focus on funds' managers behavior in stock picking, we denote the $ACC$ indicator as $ k_{F_{i},1}$. Thus, this measure differentiates among funds investing in niche assets versus those opting for more popular assets by looking at the average commonality of the constituents in the funds portfolios\footnote{We have also applied higher order measures of commonality as, for instance, ${\small k}_{F_{i},2}$ to compute investment strategies along the line described in the paper. Since results are in line with the ones presented in the paper, for sake of brevity, we exclude them from the work but they are available from authors upon request.}. In the analysis we say that a fund belongs to the low percentile of the $ACC$ distribution if the assets in its portfolio have on average low values for commonality, while the opposite occurs if its assets are popular.

\begin{table}[ht!]
\centering
\caption{\textbf{Summary Statistics.} Summary statistics of the reference quantities (Diversification, ACC and Returns) utilized throughout the paper divided into two sub-periods, namely 2004-07 and 2007-10. For each measure and for each sub-period, we report the minimum (min.), the maximum (max.), the average (mean) value of the quantity along with the standard deviation (std.), the skeweness (skew.), and the kurtosis (kurt.).} 
\scalebox{0.88}{\begin{tabular}{c|cc|cc|cc|}
        & \multicolumn{2}{c|}{\textbf{Diversification}} & \multicolumn{2}{c|}{\textbf{ACC}} & \multicolumn{2}{c|}{\textbf{Returns}} \\ \hline \hline
        & min.                  & max.                  & min.               & max.              & min.               & max.             \\
2004-07 & 1.000                 & 3027.923              & 1.187              & 207.196           & -0.022             & 0.024            \\
2007-10 & 1.000                 & 3117.364              & 1.000              & 396.776           & -0.043             & 0.042            \\ \hline \hline
        & mean                  & std.                  & mean               & std.              & mean               & std.             \\
2004-07 & 110.343               & 206.530               & 77.774             & 44.169            & 0.001              & 0.004            \\
2007-10 & 128.373               & 244.102               & 139.755            & 79.521            & 0.000              & 0.008            \\ \hline \hline
        & skew.                 & kurt.                 & skew.              & kurt.             & skew.              & kurt.            \\
2004-07 & 7.639                 & 80.404                & 0.297              & 2.062             & 0.189              & 9.696            \\
2007-10 & 6.257                 & 54.986                & 0.201              & 2.208             & 0.100              & 9.607\\
\hline
\hline
\end{tabular}}
\label{sumstat}
\end{table}

Table \ref{sumstat} shows summary statistics of the main quantities used throughout the paper, namely the diversification of a fund, its $ACC$ indicator and the gross returns. The measures are averages of the values obtained for two sub-periods. In the table we report the minimum, the maximum and the average value of each quantity along with the standard deviation, the skeweness and the kurtosis of the distribution. Notice how, during the (post)-crisis phase, all the topological measures increase both in the mean values and in their standard deviations, while the returns display higher standard deviations. On the other hand, the skeweness and the kurtosis remain approximately stable during the two sub-periods. 

\begin{figure}[h!]
\caption{\textbf{Distribution over the sample period of mean values of the ACC within the $Q10$-decile (in yellow), the $Q5$-decile (in red) and the $Q1$-decile (in blue)}. The inset reproduces the results associated with the $Q1$-decile of the ACC indicator emphasizing its dynamic over time. The associated dispersion of one standard deviation is expressed with vertical bars.}
\centering
\includegraphics[scale=0.70]{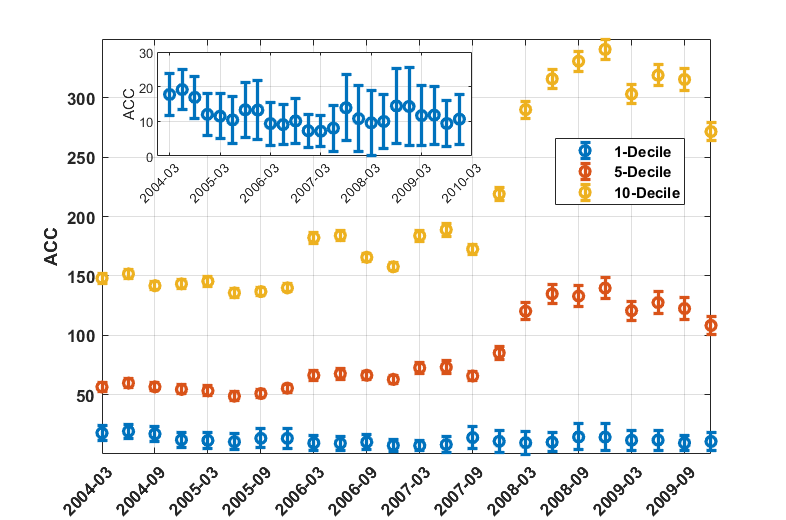}
\label{commonality_decili}
\end{figure}

Figure \ref{commonality_decili} shows the quarterly time series of mean values of the $ACC$ for funds within the $Q10$-decile (in yellow), $Q5$-decile (in red) and the $Q1$-decile (in blue) portfolios, along with the associated dispersion corresponding to one standard deviation (vertical bars). The $Q10$-decile encompasses funds with the highest $ACC$ value while the  $Q1$-decile includes those with the lowest values. We note that for the $Q10$-decile portfolio the outbreak of the financial markets of mid-2007 corresponds to an increasing trajectory in $ACC$ levels (more pronounced with respect to the increase of the $Q5$-decile), meaning that these funds increased even more their exposition to the most common/popular assets. By contrast, data for $Q1$-decile shows that investments for this class of funds remained very specialized and were not affected by the onset of the crisis.

\section{Results}
\label{results}

At the beginning of each quarter we sort funds into deciles according to different criteria: the alphas from the three- and five-factors models estimated via standard OLS procedure\footnote{We use subscripts $3f$ and $5f$ to indicate alpha computed from the three- and the five-factors model, respectively. The estimate of portfolio expected return is computed as $r_{3f} = R_{f}+\beta_{1}(R_{m}-R_{f})+\beta_{2} SMB + \beta_{3} HML + \alpha$, where the market premium ($R_{m} - R_{f}$) is enriched by factors that refer to Small minus Big capitalization (SMB) and High minus Low book-to-market ratio (HML); the five factors model adds to the previous three factors model  the profitability and the investment factors.} (\cite{fama1993common,fama2015five}), the Average Commonality Coefficient ($ACC$) and diversification ($D$) indicators, and the measure of managerial skills ($\hat{\delta^{*}}$) proposed by \cite{cohen2005judging}. The latter measure is computed as:
\begin{equation}
\hat{\delta^{*}}=W'V\hat{\alpha}
\end{equation}
where $W$ is the matrix denoting the current weight of stocks in managers portfolios and $V$ represents the fraction of stocks holding with respect to the entire funds world. 

We use nine months of daily observations as look-back period. For this reason we will refer hereinafter to the alpha sort of the first criterion as \textit{past-}$\hat{\alpha}$. Then we calculate the return of each decile portfolio over the next three months equally weighting funds in each decile. Finally, we connect quarterly decile performances 
from June 2004 to June 2010, providing also separate results for observations prior to the financial crisis of mid-2007 (namely, in the interval from June 2004 to June 2007) and for the (post)-crisis period (namely, from September 2007 to June 2010).

\begin{figure}[ht!]
\caption{\textbf{Pooled distributions of past-$\hat{\alpha}$ (x-axis) vs. ACC (y-axis) prior to the crisis (upper panels) and for the (post)-crisis period (bottom panels).} Each panel shows the scatter plot of the ACC measure vs. the past-$\hat{\alpha}$ along with the average values for both the ACC $<ACC_{3(5)f}>$ and past-$\hat{\alpha}$, $<past-\hat{\alpha}>$. In blue we report the results associated with the three-factors model($ACC_{3f}$) while the red color emphasizes the results obtained by employing the five-factors model ($ACC_{5f}$).}
\centering
\hspace{-1.5cm}\includegraphics[scale=0.6]{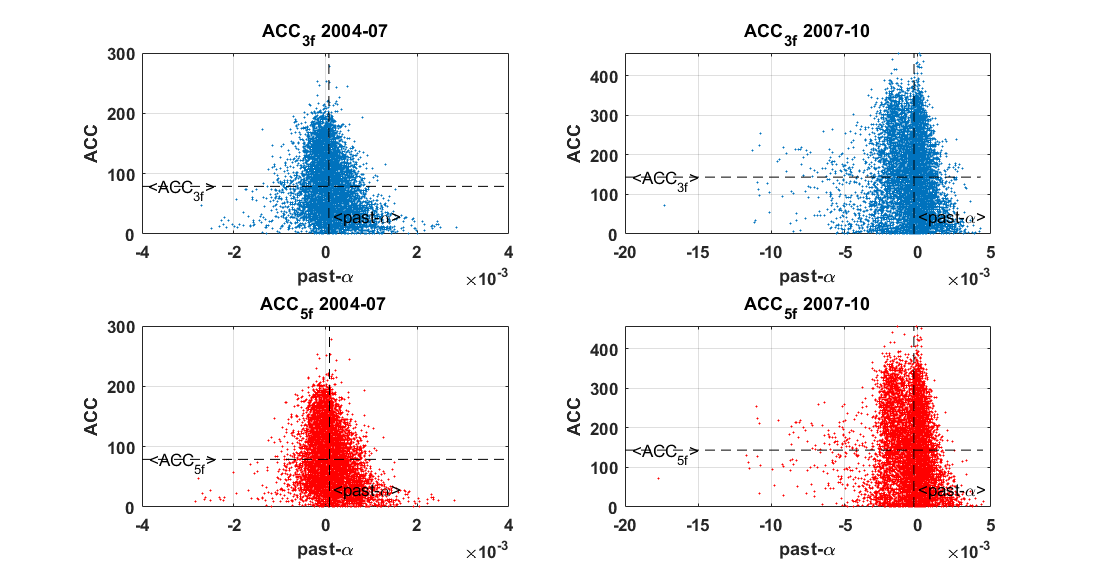}
\label{ubi_vs_alpha}
\end{figure}

Figure \ref{ubi_vs_alpha} shows the scatter plot of the past-$\hat{\alpha}$ values against the corresponding ACC value of each fund's portfolio (top panels refer to the three-factors model while bottom panels refer to the five-factors model). Prior to the crisis of mid-2007 the distribution related to the three-factors model ($ACC_{3f}$) is basically bell-shaped and centered around a value of zero for past-$\hat{\alpha}$. For the five-factors model case ($ACC_{5f}$), the distribution, besides the bell-shape configuration, displays a slightly negative average past-$\hat{\alpha}$. This configuration changes after the crisis, in time window September 2007 - June 2010, where we observe a noisier left-tail in both models. High values of ACC are thus associated with less dispersed values of past-$\hat{\alpha}$, while for funds with very low values of ACC we observe a much more variability in terms of past-$\hat{\alpha}$. Those funds which are likely to invest in popular assets are, therefore, unlikely to produce large extra-performances, while those funds investing in niche assets may get extra-performances significantly deviating from zero. Investing in less common assets deserves therefore a premium for the risk of departing from the relative performance related to the asset allocation of peer investors.

\begin{figure}[h!]
\caption{\textbf{Scatter plots of inverse ACC (G) vs. the $\hat{\delta^{*}}$ of \cite{cohen2005judging} -panel (a), vs. the Diversification measure  -panel (b) and vs. the funds' size  -panel (c).} Inside each panel we also report the results associated with the extremes of the ACC index distribution in each quarter, namely the $Q1$-decile and the $Q10$-decile. The blue points refer to the three-factors model whereas the red points correspond to the five-factors model.}
\centering
\hspace{-1cm}
\subfigure[]{\includegraphics[scale=0.25]{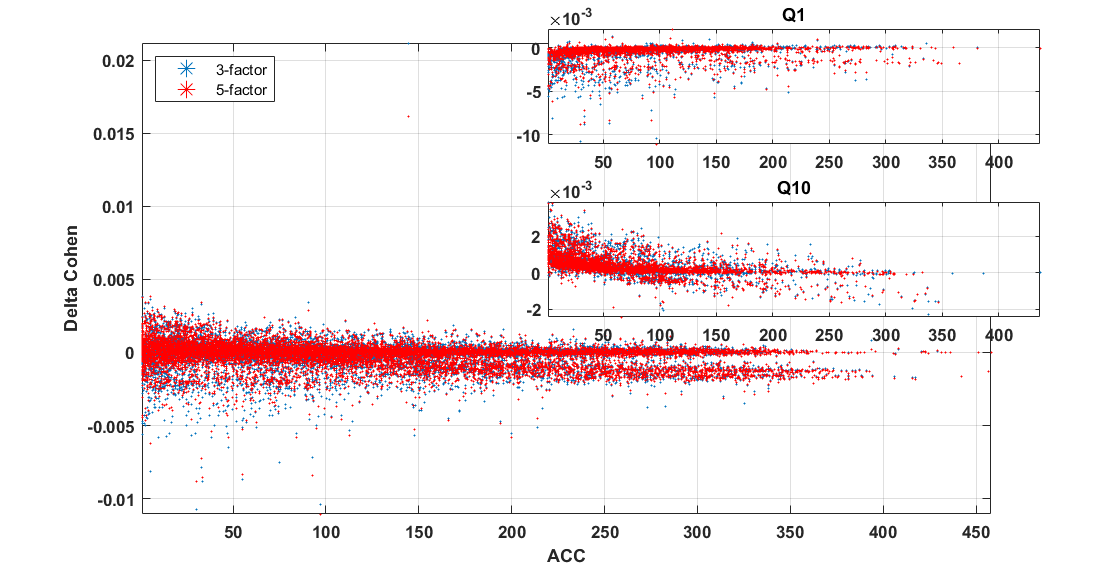}}
\subfigure[]{\includegraphics[scale=0.25]{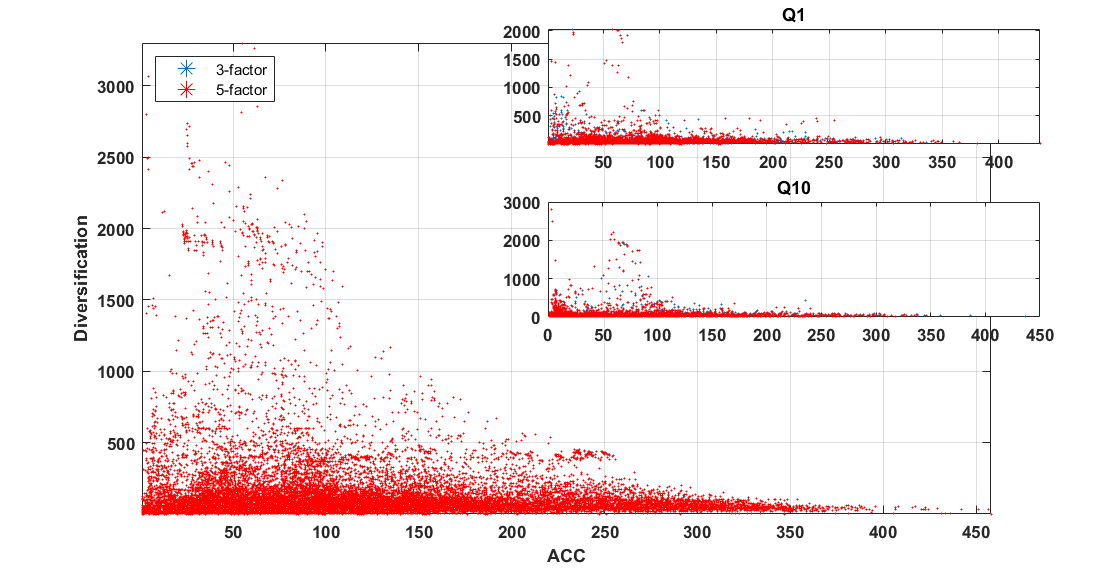}}
\subfigure[]{\includegraphics[scale=0.25]{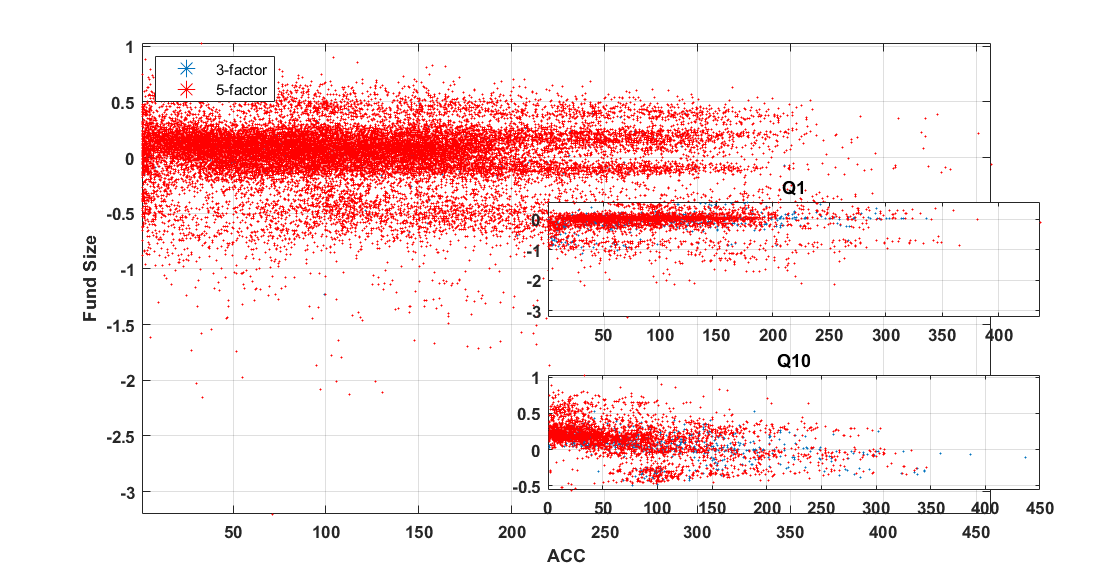}}
\label{ubi_vs_rest}
\end{figure}

Moreover, we are also interested in studying whether the information contained in the ACC indicator is not also embedded into the other measures. For this purpose, in Figure \ref{ubi_vs_rest} (a) we report the scatter plot of the ACC against the measure proposed by \cite{cohen2005judging}, in panel (b) we show the ACC vs. the diversification, and finally in panel (c) we plot the ACC against the size of the funds (computed as the sum of the market values of the constituents). Finally the inserts of each panel report the results for the $Q1$-decile and the $Q10$-decile of the ACC distribution. The figures, displaying low correlations between the variables, suggest that the information contained in the ACC measure is new and not embedded in the other variables, thus emphasizing the use of this topological measure in the horse race procedure that we will present in next subsections.

\subsection{One-way Sorting}
\label{oway}

\begin{table}[ht!]
\centering
\caption{\textbf{Sort Funds by Past Performance - Pre Crisis.} The table shows the returns of funds sorted according to various measures of past performance. The table reports the OLS estimates of decile portfolios' alphas (in percentage per year) and the corresponding absolute value of the t-statistics (in parentheses). Subscripts $3f$ and $5f$ stand for the three and five factors models used to compute alpha performances (\cite{fama1993common,fama2015five}). To compute past performance we use nine months of lookback period of daily observations on gross returns, which are determined by adding fund management fees to net returns. We then calculate the return of each decile portfolio over the next three months using daily returns series and equally weighting funds in each decile. Decile portfolios are redefined each quarter and the corresponding three months returns time series are connected across quarters to form a full sample period for each decile portfolio. With $\hat{\alpha}$ we refer to past-$\hat{\alpha}$ obtained from the three or five factors models. $ACC$ stands for the Average Commonality Coefficient property, while letter $D$ refers to the diversification. Finally, with $\hat{\delta^{*}}$ we refer to the delta measure of managerial skill introduced by \cite{cohen2005judging}. \textit{top-bottom} is the portfolio obtained investing long in the best-performer decile and short funds in the worst-performer decile. The sample period is June 2004 to June 2007.}
\scalebox{.7}{\begin{tabular}{cccccccccccc}
\hline
\hline
\multicolumn{1}{l}{}     & Q1      & Q2      & Q3      & Q4      & Q5      & Q6      & Q7      & Q8      & Q9      & Q10 &  \textit{top-bottom}   \\ \hline \hline
 &       &     &       &       &       &       &       &       &       &  &     \\

    \multicolumn{1}{l}{$\hat{\alpha}_{5f}$} &     2.74  & 1.51  & 1.56  & 1.96  & 2.21  & 1.96  & 3.19  & 3.38  & 4.01  & 11.51 & 8.77 \\
 &   (1.31)  & (1.59)  & (1.87)  & (2.22)  & (2.75)  & (2.22)  & (3.21)  & (2.73)  & (2.63)  & (2.83)  & (2.05) \\
       &         &         &         &         &         &         &         &         &         &         &         \\

    \multicolumn{1}{l}{$\hat{\alpha}_{3f}$} & 0.87  & 0.64  & 0.19  & 1.82  & 1.83  & 2.25  & 2.63  & 3.66  & 4.94  & 11.58 & 10.71 \\
 &   (0.53)  & (0.70)  & (0.24)  & (2.40)  & (2.37)  & (2.69)  & (2.93) & (2.93)  & (3.26)  & (2.84)  & (2.62) \\
       &         &         &         &         &         &         &         &         &         &         &         \\

    \multicolumn{1}{l}{$ACC_{5f}$} &     11.80 & 7.53  & 3.95  & 3.07  & 4.99  & 3.04  & 3.03  & 1.91  & 0.83  & -0.05 & 11.85 \\
 &   (3.14)  & (3.05)  & (3.30)  & (2.31)  & (3.46)  & (2.40)  & (2.82)  & (2.59)  & (1.32)  & (0.10) & (3.23) \\

     &       &     &       &       &       &       &       &       &       &  &     \\

    \multicolumn{1}{l}{$ACC_{3f}$}  & 11.22 & 7.13  & 3.64  & 2.70  & 4.58  & 2.76  & 2.82  & 1.79  & 0.72  & -0.13 & 11.36 \\
  &  (3.00)  & (2.91)  & (3.16)  & (2.11)  & (3.26)  & (2.23)  & (2.68)  & (2.48)  & (1.19)  & (0.26) & (3.12) \\
     &       &     &       &       &       &       &       &       &       &  &     \\

    \multicolumn{1}{l}{$D_{5f}$} &     10.85 & 6.80  & 4.35  & 3.66  & 3.11  & 2.75  & 1.91  & 2.42  & 1.90  & 0.57  & 10.28 \\
  &  (2.68)  & (2.72)  & (3.35)  & (3.32)  & (3.45)  & (2.88)  & (2.04)  & (2.84)  & (2.23)  & (0.98)  & (2.67) \\
     &       &     &       &       &       &       &       &       &       &  &     \\

    \multicolumn{1}{l}{$D_{3f}$} &    10.65 & 6.33  & 4.00  & 3.26  & 2.80  & 2.52  & 1.63  & 2.17  & 1.68  & 0.45  & 10.20 \\
  &  (2.62)  & (2.55)  & (3.14)  & (3.06)  & (3.21)  & (2.74)  & (1.79)  & (2.64)  & (2.02) & (0.80)  & (2.64) \\
     &       &     &       &       &       &       &       &       &       &  &     \\

    \multicolumn{1}{l}{$\hat{\delta^{*}}_{5f}$} &     4.44  & 1.98  & 1.37  & 1.23  & 1.29  & 1.61  & 2.15  & 2.34  & 4.30  & 12.40 & 7.96 \\
 &   (2.08)  & (1.50)  & (1.42)  & (1.65)  & (1.66)  & (1.84)  & (2.11)  & (1.94)  & (2.83)  & (2.89)  & (1.83) \\
   \\

    \multicolumn{1}{l}{$\hat{\delta^{*}}_{3f}$} &    2.51  & 1.12  & 0.87  & 1.09  & 1.02  & 1.18  & 1.57  & 2.36  & 4.59  & 12.89 & 10.39 \\
   & (1.39)  & (1.10)  & (1.15)  & (1.56)  & (1.43)  & (1.44)  & (1.64)  & (1.97)  & (3.03)  & (2.94)  & (2.47) \\
     &       &     &       &       &       &       &       &       &       &  &     \\

 \hline
 \hline
\end{tabular}}
\label{tab:alpha_pre}
\end{table}

\begin{table}[ht!]
\centering
\caption{\textbf{Sort Funds by Past Performance - Crisis.} The table shows the returns of funds sorted according to various measures of past performance. The table reports the OLS estimates of decile portfolios' alphas (in percentage per year) and the corresponding absolute value of the t-statistics (in parentheses). Subscripts $3f$ and $5f$ stand for the three and five factors models used to compute alpha performances (\cite{fama1993common,fama2015five}). To compute past performance we use nine months of lookback period of daily observations on gross returns, which are determined by adding fund management fees to net returns. We then calculate the return of each decile portfolio over the next three months using daily returns series and equally weighting funds in each decile. Decile portfolios are redefined each quarter and the corresponding three months returns time series are connected across quarters to form a full sample period for each decile portfolio. With $\hat{\alpha}$ we refer to past-$\hat{\alpha}$ obtained from the three or five factors models. $ACC$ stands for the Average Commonality Coefficient property, while letter $D$ refers to the diversification. Finally, with $\hat{\delta^{*}}$ we refer to the delta measure of managerial skill introduced by \cite{cohen2005judging}. \textit{top-bottom} is the portfolio obtained investing long in the best-performer decile and short funds in the worst-performer decile. The sample period is September 2007 to June 2010.}
\scalebox{.7}{\begin{tabular}{cccccccccccc}
\hline
\hline
\multicolumn{1}{l}{}     & Q1      & Q2      & Q3      & Q4      & Q5      & Q6      & Q7      & Q8      & Q9      & Q10 &  \textit{top-bottom}   \\ \hline \hline
 &       &     &       &       &       &       &       &       &       &  &     \\

    \multicolumn{1}{l}{$\hat{\alpha}_{5f}$} &         -14.74 & -5.17 & -7.34 & -5.37 & -5.70 & -4.15 & -5.14 & -5.27 & -4.56 & -5.84 & 8.90 \\
 &   (2.11) & (1.04) & (1.82) & (1.68) & (1.76) & (1.27) & (1.41) & (1.51) & (1.27) & (0.88) & (1.07) \\

       &         &         &         &         &         &         &         &         &         &         &         \\

    \multicolumn{1}{l}{$\hat{\alpha}_{3f}$} &     -13.10 & -7.05 & -7.93 & -5.38 & -5.27 & -4.24 & -4.79 & -5.38 & -3.69 & -3.34 & 9.76 \\
  &  (2.06) & (1.34) & (1.91) & (1.59) & (1.58) & (1.27) & (1.37) & (1.49) & (1.22) & (0.65) & (1.44) \\

       &         &         &         &         &         &         &         &         &         &         &         \\

    \multicolumn{1}{l}{$ACC_{5f}$} &         -7.51 & -5.41 & -4.02 & -4.23 & -5.61 & -5.97 & -4.91 & -7.56 & -7.54 & -6.59 & -0.92 \\
 &   (1.04) & (0.97) & (1.41) & (1.38) & (1.58) & (1.52) & (1.33) & (1.95) & (1.95) & (1.81) & (0.13) \\

     &       &     &       &       &       &       &       &       &       &  &     \\

    \multicolumn{1}{l}{$ACC_{3f}$}  &     -6.11 & -4.61 & -4.07 & -3.91 & -5.20 & -5.76 & -4.97 & -7.08 & -7.42 & -7.09 & 0.98 \\
 &   (0.86) & (0.85) & (1.52) & (1.35) & (1.51) & (1.50) & (1.37) & (1.84) & (1.92) & (1.94) & (0.14) \\

     &       &     &       &       &       &       &       &       &       &  &     \\

    \multicolumn{1}{l}{$D_{5f}$} &     -7.24 & -6.09 & -5.69 & -5.34 & -6.18 & -5.45 & -7.50 & -5.95 & -5.52 & -5.93 & -1.31 \\
 &   (1.10) & (1.24) & (1.64) & (1.52) & (1.78) & (1.50) & (2.11) & (1.76) & (1.70) & (1.89) & (0.26) \\
     &       &     &       &       &       &       &       &       &       &  &     \\

    \multicolumn{1}{l}{$D_{3f}$} &        -6.53 & -5.19 & -5.40 & -5.08 & -5.82 & -5.47 & -7.10 & -5.81 & -5.46 & -6.03 & -0.50 \\
 &   (0.99) & (1.07) & (1.58) & (1.47) & (1.72) & (1.54) & (2.08) & (1.77) & (1.72) & (1.96) & (0.10) \\
     &       &     &       &       &       &       &       &       &       &  &     \\

    \multicolumn{1}{l}{$\hat{\delta^{*}}_{5f}$} &      -11.01 & -8.12 & -6.02 & -6.47 & -5.82 & -5.03 & -4.70 & -5.14 & -4.98 & -6.66 & 4.35 \\
 &   (1.47) & (1.56) & (1.58) & (1.93) & (1.57) & (1.45) & (1.32) & (1.46) & (1.43) & (0.99) & (0.48) \\
    &       &     &       &       &       &       &       &       &       &  &     \\

    \multicolumn{1}{l}{$\hat{\delta^{*}}_{3f}$} &       -10.93 & -7.56 & -6.50 & -7.37 & -6.93 & -4.99 & -3.92 & -5.49 & -3.31 & -3.76 & 7.17 \\
 &   (1.47) & (1.43) & (1.62) & (2.02) & (1.77) & (1.34) & (1.15) & (1.71) & (1.02) & (0.71) & (0.91) \\
    &       &     &       &       &       &       &       &       &       &  &     \\

 \hline
 \hline
\end{tabular}}
\label{tab:alpha_post}
\end{table}

\begin{table}[ht!]
\centering
\caption{\textbf{Sort Funds by Past Performance.} The table shows the returns of funds sorted according to various measures of past performance. The table reports the OLS estimates of decile portfolios' alphas (in percentage per year) and the corresponding absolute value of the t-statistics (in parentheses). Subscripts $3f$ and $5f$ stand for the three and five factors models used to compute alpha performances (\cite{fama1993common,fama2015five}). To compute past performance we use nine months of lookback period of daily observations on gross returns, which are determined by adding fund management fees to net returns. We then calculate the return of each decile portfolio over the next three months using daily returns series and equally weighting funds in each decile. Decile portfolios are redefined each quarter and the corresponding three months returns time series are connected across quarters to form a full sample period for each decile portfolio. With $\hat{\alpha}$ we refer to past-$\hat{\alpha}$ obtained from the three or five factors models. $ACC$ stands for the Average Commonality Coefficient property, while letter $D$ refers to the diversification. Finally, with $\hat{\delta^{*}}$ we refer to the delta measure of managerial skill introduced by \cite{cohen2005judging}. \textit{top-bottom} is the portfolio obtained investing long in the best-performer decile and short funds in the worst-performer decile. The sample period is June 2004 to June 2010.}
\scalebox{.7}{\begin{tabular}{cccccccccccc}
\hline
\hline
\multicolumn{1}{l}{}     & Q1      & Q2      & Q3      & Q4      & Q5      & Q6      & Q7      & Q8      & Q9      & Q10 &  \textit{top-bottom}   \\ \hline \hline
 &       &     &       &       &       &       &       &       &       &  &     \\

    \multicolumn{1}{l}{$\hat{\alpha}_{5f}$} &              -5.28 & -1.38 & -2.31 & -1.25 & -1.21 & -0.46 & -0.31 & -0.33 & 0.39  & 3.65  & 8.93 \\
& (1.52) & (0.58) & (1.20) & (0.80) & (0.78) & (0.29) & (0.18) & (0.19) & (0.21)  & (0.97)  & (1.94) \\
&       &       &       &       &       &       &       &       &       &       &  \\
                    
 \multicolumn{1}{l}{$\hat{\alpha}_{3f}$}    &    -5.14 & -2.57 & -3.28 & -1.25 & -1.16 & -0.43 & -0.51 & -0.12 & 1.37  & 5.12  & 10.25 \\
& (1.68) & (1.04) & (1.67) & (0.77) & (0.73) & (0.27) & (0.30) & (0.07) & (0.84)  & (1.57)  & (2.64) \\
&       &       &       &       &       &       &       &       &       &       &  \\

 \multicolumn{1}{l}{$ACC_{5f}$}   &  3.62  & 1.96  & 0.24  & -0.10 & 0.21  & -0.82 & -0.42 & -2.40 & -2.85 & -2.89 & 6.51 \\
& (0.92)  & (0.68)  & (0.16)  & (0.06) & (0.12)  & (0.42) & (0.23) & (1.30) & (1.56) & (1.66) & (1.67) \\
&       &       &       &       &       &       &       &       &       &       &  \\
          
 \multicolumn{1}{l}{$ACC_{3f}$}        & 4.00  & 1.95  & 0.16  & -0.18 & 0.20  & -0.88 & -0.41 & -2.09 & -2.60 & -2.88 & 6.88 \\
& (1.03)  & (0.68)  & (0.11)  & (0.12) & (0.12)  & (0.47) & (0.23) & (1.15) & (1.43) & (1.65) & (1.77) \\
&       &       &       &       &       &       &       &       &       &       &  \\

 \multicolumn{1}{l}{$D_{5f}$}         & 2.92  & 1.22  & -0.02 & -0.30 & -0.95 & -0.97 & -2.20 & -1.27 & -1.30 & -2.14 & 5.06 \\
& (0.78)  & (0.46)  & (0.01) & (0.18) & (0.57) & (0.55) & (1.28) & (0.77) & (0.82) & (1.43) & (1.61) \\
&       &       &       &       &       &       &       &       &       &       &  \\          
          
 \multicolumn{1}{l}{$D_{3f}$}        & 2.85  & 1.42  & -0.04 & -0.25 & -0.80 & -0.78 & -2.14 & -1.12 & -1.24 & -2.24 & 5.09 \\
& (0.76)  & (0.54)  & (0.02) & (0.15) & (0.49) & (0.45) & (1.29) & (0.70) & (0.81) & (1.53) & (1.62) \\
&       &       &       &       &       &       &       &       &       &       &  \\

 \multicolumn{1}{l}{$\hat{\delta^{*}}_{5f}$}       & -2.19 & -2.41 & -2.10 & -2.32 & -1.73 & -1.15 & -0.66 & -0.67 & 0.34  & 3.59  & 5.78 \\
& (0.59) & (0.95) & (1.13) & (1.43) & (0.97) & (0.69) & (0.38) & (0.38) & (0.19)  & (0.91)  & (1.17) \\
&       &       &       &       &       &       &       &       &       &       &  \\
          
 \multicolumn{1}{l}{$\hat{\delta^{*}}_{3f}$}            & -2.99 & -2.46 & -2.31 & -2.62 & -2.31 & -1.26 & -0.56 & -0.88 & 1.26  & 5.28  & 8.28 \\
& (0.84) & (0.98) & (1.21) & (1.52) & (1.25) & (0.71) & (0.34) & (0.54) & (0.73)  & (1.54)  & (1.88) \\
&       &     &       &       &       &       &       &       &       &  &     \\

 \hline
 \hline
\end{tabular}}
\label{tab:alpha_all}
\end{table}

In this Section we report the results of the comparison of the alphas for each decile according to different measures of past performance, with the aim of identifying the one which gives highest profitability before and after the crisis.
Table \ref{tab:alpha_pre} shows the annualized post-ranking alphas for each decile portfolio (sorted from the lowest Q1 to the highest Q10) during the pre-crisis period, calculated using both the three- and five- factors models. It also reports the performance of the portfolio long in the top best performer decile and short in the bottom worst performer decile (namely, \textit{top-bottom} portfolio\footnote{For the topological properties \textit{ACC} and \textit{diversification} the top best performer decile refers to Q1, while it is Q10 for past-$\hat{\alpha}$ and the $\hat{\delta^{*}}$ indicator of \cite{cohen2005judging}.}). The performance of the \textit{top-bottom} portfolio built according to the past-$\hat{\alpha}$ sort produces consistent (about [8.77; 10.71]) and significant (t-statistics [2.05; 2.62]) annual returns, thus supporting the view of predictability of funds' returns based on past-$\hat{\alpha}$ performances. By using, as ranking measure, the ACC or the diversification property we get significant and even higher \textit{top-bottom} annual extra-performances (about [11.85; 11.36] per cent for the ACC and [10.28; 10.20] per cent for diversification). In addition, as alternative indicator to judge managerial skills we report the performances related to the sort of past performances based on the measure proposed by \cite{cohen2005judging}, obtaining similar results as those related to the other ranking criteria (about [7.96; 10.39] per cent). These results are significant not only economically but also statistically, supporting the use of these sort criteria to build portfolios. Hence, all the measures presented in Table \ref{tab:alpha_pre} seem capable of generating future extra-performances when combined in a \textit{top-bottom} portfolio strategy.

The impact of 2007-08 financial crisis deteriorated market performances and heavily impacted the mutual fund sector. Despite the negative outcomes occurred along the interval 2007-2010, annualized extra-performances for \textit{top-bottom} portfolios shown in Table \ref{tab:alpha_post} are still positive when using past-$\hat{\alpha}$ or $\hat{\delta^{*}}$ as sorting criteria, although poorly statistically significant. Conversely, our topological indicators seem not able to determine positive results, presenting also a not clear monotonic pattern along the decile portfolios. Thus, the crisis of 2007-2008 undermined the relationship between performances and the topological properties of the portfolios that, instead, emerged as a complementary source of information prior to the crisis.

More generally, results for the entire sample period show positive and consistent \textit{top-bottom} extra-performances (in a range from about 5 to 10 per cent) as reported in Table \ref{tab:alpha_all}. Interestingly, \textit{top-bottom} portfolios obtained using both past-$\hat{\alpha}$ and $\hat{\delta^{*}}$ sorting criteria appear less affected by the onset of the financial crisis, while sorting according to the topological indicator reflects the changes occurred across the crisis and that perturbed the relationship with performances as discussed above. Finally, for each sorting criteria we confirm that results for the overall sample period indicate a clear monotonic pattern in the way the corresponding horse-race strategy generates extra-performances across the decile portfolios. Findings are finally robust to the choice to utilize the three- or five-factors models to measure alphas, being very similar in terms of resulting extra-performances and significance levels in each period.

To limit potential issues due to the presence of outliers in the distribution of ACC, we finally drop those funds in the tails corresponding to both the top and bottom five per cent of the ACC distribution in each decile portfolio. Table \ref{tab:alpha_pre_core} shows this \textit{Core} ACC case, confirming that previous findings hold even for more cohesive decile partitions.

\begin{table}[ht!]
	\centering
	\caption{\textbf{Sort Funds by Past Performance - Core case.} The table exhibits the one-way sort performances for decile portfolios based on ACC. For each decile portfolio we drop those funds in the tails corresponding to both the top and bottom 5 per cent of the ACC distribution in each decile portfolio. The table reports the OLS estimates of decile portfolios' alphas (in percentage per year) and the corresponding absolute value of the t-statistics (in parentheses). Subscripts $3f$ and $5f$ stand for the three and five factor models used to compute alpha performances (\cite{fama1993common,fama2015five}). To compute past performance we use nine months of lookback period of daily observations on gross returns, which are determined by adding fund management fees to net returns. We then calculate the return of each decile portfolio over the next three months using daily returns series and equally weighting funds in each decile. Decile portfolios are redefined each quarter and the corresponding three months returns time series are connected across quarters to form a full sample period for each decile portfolio. \textit{top-bottom} is the portfolio obtained investing long in the best-performer decile and short funds in the worst-performer decile. Panel A stands for the interval from June 2004 to June 2007, Panel B refers to the period from September 2007 to June 2010, and Panel C from June 2004 to June 2010.}
	\scalebox{.7}{\begin{tabular}{ccccccccccccc}
			\hline
            \hline
			\multicolumn{1}{l}{}     & Q1      & Q2      & Q3      & Q4      & Q5      & Q6      & Q7      & Q8      & Q9      & Q10 &  \textit{top-bottom}   \\ 
            \hline
            \hline
			
		\multicolumn{1}{l}{Panel A: 2004-2007} &       &       &       &       &       &       &       &       &       &  \\
			
		\multicolumn{1}{l}{$ACC_{5f}$} &   12.24 & 7.39  & 4.00  & 2.95  & 5.01  & 2.90  & 2.83  & 1.90  & 0.84  & 0.05  & 12.20 \\
		& (3.13)  & (3.01)  & (3.31)  & (2.22)  & (3.48)  & (2.28)  & (2.62)  & (2.58)  & (1.30)  & (0.08)  & (3.19) \\
	\multicolumn{1}{l}{$ACC_{3f}$}	& 11.64 & 7.00  & 3.69  & 2.60  & 4.60  & 2.63  & 2.63  & 1.77  & 0.74  & -0.06 & 11.70 \\
		& (2.99)  & (2.87)  & (3.17)  & (2.02)  & (3.29)  & (2.11)  & (2.48)  & (2.47)  & (1.18)  & (0.11) & (3.08) \\
		&       &       &       &       &       &       &       &       &       &  \\
		\multicolumn{1}{l}{Panel B: 2007-2010} &       &       &       &       &       &       &       &       &       &  \\
		
		\multicolumn{1}{l}{$ACC_{5f}$} &-7.44 & -5.53 & -4.21 & -4.30 & -5.53 & -6.23 & -5.04 & -7.39 & -7.49 & -6.39 & -1.05 \\
		& (1.01) & (1.00) & (1.48) & (1.41) & (1.53) & (1.57) & (1.33) & (1.91) & (1.92) & (1.76) & (0.14) \\
		\multicolumn{1}{l}{$ACC_{3f}$} &-6.03 & -4.74 & -4.29 & -3.93 & -5.10 & -6.02 & -5.15 & -6.94 & -7.38 & -6.87 & 0.84 \\
		& (0.82) & (0.87) & (1.61) & (1.37) & (1.45) & (1.56) & (1.39) & (1.81) & (1.89) & (1.89) & (0.11) \\
		&       &       &       &       &       &       &       &       &       &  \\
		\multicolumn{1}{l}{Panel C: 2004-2010} &       &       &       &       &       &       &       &       &       &  \\
		
		\multicolumn{1}{l}{$ACC_{5f}$}  &	3.89  & 1.85  & 0.18  & -0.19 & 0.27  & -1.04 & -0.57 & -2.32 & -2.82 & -2.72 & 6.61 \\
	&	(0.96)  & (0.64)  & (0.12)  & (0.12) & (0.15)  & (0.53) & (0.31) & (1.26) & (1.52) & (1.57) & (1.64) \\
		\multicolumn{1}{l}{$ACC_{3f}$}	&   4.30  & 1.82  & 0.08  & -0.25 & 0.28  & -1.09 & -0.60 & -2.03 & -2.55 & -2.72 & 7.02 \\
	&	(1.07)  & (0.64)  & (0.06)  & (0.16) & (0.16)  & (0.57) & (0.33) & (1.12) & (1.38) & (1.57) & (1.75) \\
    \\
		
			\hline
			\hline
	\end{tabular}}
	\label{tab:alpha_pre_core}
\end{table}

\subsection{Inspecting the Mechanism behind the Performance of the \textit{ACC} Indicator}

The previous analysis highlights a positive correlation between fund performance and the level of specialization of its portfolio. Funds with a low ACC portfolios (high specialization) seem to gain higher extra-performances with respect to funds that invest in more popular assets. This can be due to the fact that those specialized funds may be more informed and more able to extract profits from this information. Nevertheless, the 2007 financial crisis modified the relationships between funds and constituents, thus also the connection between the ACC index and the past-$\hat{\alpha}$. 

We investigate with a higher time resolution the effects of the crisis on funds' performances and the relationship between performance and specialization. Previous results are obtained by dividing the sample into two sub-periods, from 2004 to 2007 and from 2007 to 2010. This aggregation prevents to focus specifically on the crisis period, therefore, in this subsection, we show the horse-race results on a more granular time scale that is quarter by quarter. Since our main goal is to assess whether the ACC measure reveals some information on funds' performance, we focus on the one-way sort. As before we employ nine months as look-back period but the alpha values are here computed for two portfolios only, namely the high-specialization (low ACC) portfolio and the low-specialization portfolio (high ACC). As a reference threshold to divide the funds into these two samples, we apply the median value of the ACC\footnote{This choice prevents estimates with few data points. Results are qualitatively similar to those obtained using tertiles for the ACC distribution.}.

\begin{figure}[ht!]
\caption{\textbf{Annualized quarterly extra-performance ($\hat{\alpha}$) for portfolios composed by low-ACC (red) funds and high-ACC (blue) funds}. Each plot shows the annualized $\hat{\alpha}$ computed at each quarter for the high vs. low ACC portfolios together with the appropriate standard deviation (dashed lines). The left panel reports the results obtained with the three-factors model while the right panel encompasses the extra-returns computed with the five-factors model.}
\centering
\includegraphics[scale=0.50]{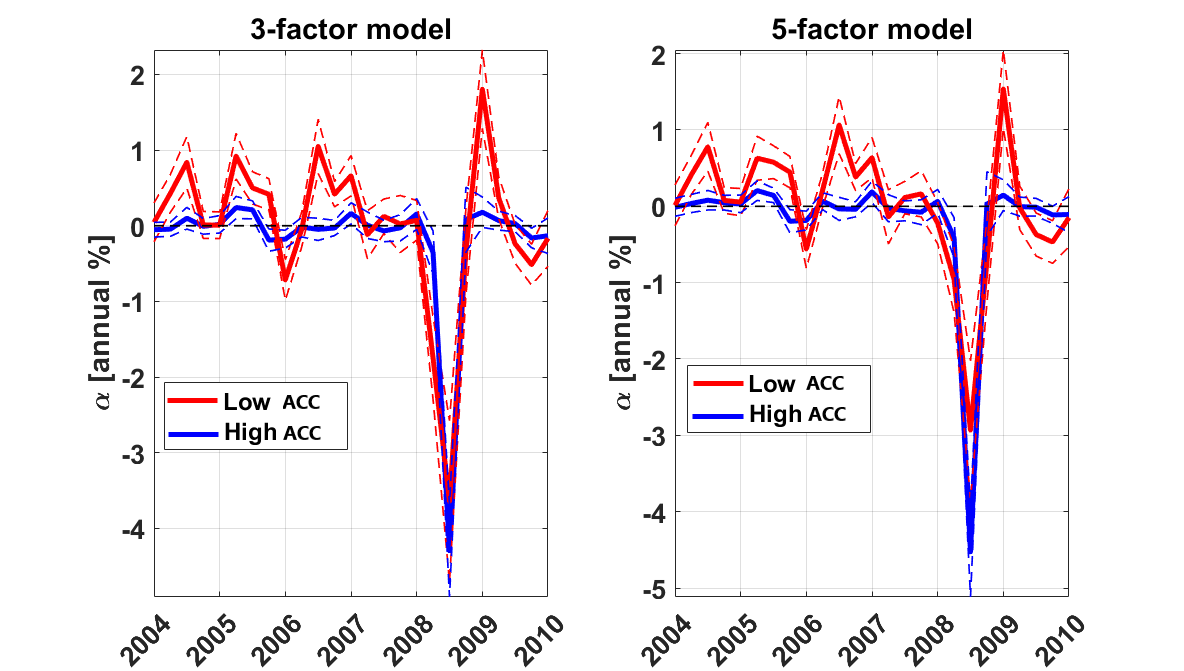}
\label{commonality_HL}
\end{figure} 

Figure \ref{commonality_HL} reports the annualized quarterly extra-performances ($\hat{\alpha}$) obtained by employing the three-factors model (left) and the five-factors model (right). From the figure clearly emerges that funds with a relative low value of $ACC$ on average perform better than funds investing in more popular assets. The red line, indicating the $\hat{\alpha}$ value for a portfolio composed by low-$ACC$ funds, is almost always positive, thus reinforcing the finding that funds investing in market niches are more likely to be informed and able to extract profit from these exposures. The blue line, on the other hand, suggests that funds investing in more common assets obtain virtually null $\hat{\alpha}$, meaning that are not able to beat systematically the market.

The years of the global financial crisis deteriorate the performances of all the funds irrespective from their $ACC$ values. Nevertheless, funds with a high $ACC$ value seem to be more affected by the crisis in both the three-factors and five-factors models. The low $ACC$ of funds' portfolios arises as an important topological property, from an investor perspective, also during crisis phases. Indeed, specialized funds investing in niche assets, despite suffering for the systemic impact of the crisis of mid-2007, seem to be less affected by the second round of the crisis in which fire sales deteriorate most the market prices of commonly holding assets.

\subsection{Double Sorts}
\label{ds}

To better understand the results of the above horse-race, 
we perform a more in depth analysis within each decile. In particular, we are interested in understanding whether the topological properties of the funds have information not contained in the alpha measures and that can be, therefore, exploited to forecast funds' performances. We focus on the use of the more sophisticated topological measure, i.e. the $ACC$ measure, by employing a double sort between past-$\hat{\alpha}$ quintiles and, within these quintiles, by further splitting funds in quintiles according to the $ACC$ levels of the portfolios. The resulting 5x5 portfolios are then mapped in time, with quarterly rebalancing, to study the distribution of alpha performances as a function of this topological property but given the same quintile level of past-$\hat{\alpha}$ performance in the first sort. Tables \ref{tab:doublesorting_pre}-\ref{tab:double_all} report, for different time windows, the resulting alpha performances for the 5x5 portfolios as well as for the \textit{top-bottom} portfolios that buy funds with low $ACC$ values and short funds with high values of $ACC$ within a given past-$\hat{\alpha}$ quintile. Finally, the portfolio denoted as \textit{Avg} invests equally in each of the five quintile portfolios by row, thus representing our cleanest measure of whether the $ACC$ index contains additional information. To provide robustness for our results, panels B in Tables \ref{tab:doublesorting_pre}-\ref{tab:double_all} present the results related to the \textit{Core} case for $ACC$ where we basically drop those funds in the tails corresponding to both the top and bottom 5 per cent of the $ACC$ distribution in each 5x5 portfolio. 

Empirical findings in Table \ref{tab:doublesorting_pre} show that those funds belonging to the top past-$\hat{\alpha}$ quintile (i.e., Q5) are not only able to generate persistent extra-performances than those in the bottom past-$\hat{\alpha}$ quintile (i.e., Q1), as already seen in subsection \ref{oway}, but that, focusing within each of these past-$\hat{\alpha}$ quintiles, we get different distributions of performances based on the level of the $ACC$ index. Interestingly, the average difference prior to the crisis between the \textit{top-bottom} quintiles ranked by $ACC$ is above 6 per cent, being significant both economically and statistically. These extra-performances suggest, therefore, that the ACC index contains information above and beyond past-$\hat{\alpha}$ sort that can be exploited to forecast funds' returns. 

The $ACC$ property seems to have a substantial impact on the extra-performances of the top past-$\hat{\alpha}$ quintile. In fact, the double sorts procedure indicates that, among funds with higher past-$\hat{\alpha}$ (i.e., Q5), those with more specialized portfolios (low $ACC$) (i.e., ACC1) are more likely to obtain higher extra-returns than those with portfolios characterized by more popular assets (i.e., ACC5). In particular, annualized extra-performances for portfolio Q5ACC1 is about [17.10; 17.98] per cent, while for Q5ACC5 is about [3.73; 5.51] per cent, with decreasing pattern in the middle of the $ACC$ distribution within Q5. Hence, very skilled managers, namely those with high past-$\hat{\alpha}$, and with niche investment exposures, namely investing in assets not very common across other portfolios, are more prone to produce substantial positive extra-performances. This result suggests that managerial skills in detecting and picking assets are practically more effective especially for those managers with better past performances (i.e., Q5ACC1). By contrast, those funds in the bottom-alpha quintile (i.e., Q1) reach lower performances and do not show a clear relationship with the $ACC$ property of their portfolios. For the latter, fund managers investing in niche or popular assets do not seem to be in general really informative, while among skilled managers, those investing in less common assets are likely to generate better future returns, at least prior to the crisis of mid-2007.

Conversely, results for the crisis period (i.e., 2007-2010) depict the $ACC$ property as less able to add valuable information in the construction of better performing portfolios (see Table \ref{tab:doublesorting_post}). Once having previously partitioned the sample according to past-$\hat{\alpha}$, still lower values for the $ACC$ index seem to generate better results than for higher ones, but results are in general not statistically significant. After the outbreak of financial markets, fund managers seem less able to gain from investing in niche vs. popular assets as instead we observed from allocations prior to the crisis. The systemic crisis affecting the global economy in mid-2007 is likely to have made financial markets much more correlated than in the previous years, thus reducing the attitude of fund managers to deviate from common investment behaviors in the pursuit of controlling for relative performance within the mutual fund industry. Despite the fact that results reported in Tables \ref{tab:doublesorting_pre}-\ref{tab:doublesorting_post} are weakly significant from a statistical view point; the difference of performances between ACC1 and ACC5 is always in favor of ACC1 (with the only exception of Q1 in Table \ref{tab:doublesorting_post}).

Finally, in Table \ref{tab:double_all} we report the extra-performances obtained for the entire sample period 2004-2010. Here, we confirm the role of $ACC$ value in discriminating portfolios' performances, with lower values for the topological indicator signaling better investment allocations. This, in turn, supports the use of our proposed indicator as a complementary criterion than past-$\hat{\alpha}$ for building profitable investment strategies on a longer holding period, when markets experienced a boom and burst financial cycle. Furthermore, we note that results for the \textit{Core} case confirm our findings on the profitability of the \textit{top-bottom} investment strategy.

\begin{table}[ht!]
	\centering
	\caption{\textbf{Double Sorts Funds by Past Performances and $ACC$ - Pre Crisis.} The table shows the resulting 5x5 portfolios' alphas obtained by sorting funds in quintiles firstly by past-$\hat{\alpha}$ performances and then by $ACC$. Panel B uses the core $ACC$ in which we drop those funds in the tails corresponding to both the top and bottom 5 per cent of the $ACC$ distribution in each 5x5 portfolio. The table reports the OLS estimates of each 5x5 portfolio's alpha (in percentage per year) and the corresponding absolute value of the t-statistics (in parentheses). Subscripts $3f$ and $5f$ stand for the three and five factors models used to compute alpha performances (\cite{fama1993common,fama2015five}). To compute past performance we use nine months of lookback period of daily observations on gross returns, which are determined by adding fund management fees to net returns. We then calculate the return of each quintile portfolio over the next three months using daily returns series and equally weighting funds in each 5x5 portfolio. 5x5 portfolios are redefined each quarter and the corresponding three months returns time series are connected across quarters to form a full sample period for each 5x5 portfolio. \textit{top-bottom} is the portfolio obtained investing long in funds belonging to the best-performer quintile portfolio and short funds in the worst-performer quintile, within the same quintile determined in the first sort. Finally, the portfolio denoted as \textit{Avg} invests equally in each of the five \textit{top-bottom} portfolios. The sample period is June 2004 to June 2007.}
	\scalebox{0.60}{  \begin{tabular}{lccccccp{0.2cm}lcccccc}
			\hline
			\hline
			\\
			&       &     &  \multicolumn{11}{c}{Panel A: Sort funds by past-$\hat{\alpha}$ and then by $ACC$}   \\            
			&     &     &     &     &     &   &       &  &     &     &     &    &   &  \\
			$\hat{\alpha}_{3f}$ $\rightarrow$ & Q1    & Q2    & Q3    & Q4    & Q5    & Avg  &       & $\hat{\alpha}_{5f}$ $\rightarrow$ & Q1    & Q2    & Q3    & Q4    & Q5    & Avg \\
			$\downarrow$ $ACC$ &     &     &     &     &     &   &       & $\downarrow$ $ACC$  &     &     &     &     &     &  \\
			
			\\
			\hline
			\\
			ACC1    &     5.26  & 4.81  & 4.66  & 5.78  & 17.10 & 7.52  &       &  ACC1     & 4.37  & 2.81  & 4.37  & 5.92  & 17.98 & 7.09 \\
			&  (1.56)  & (2.29)  & (2.27)  & (2.81)  & (3.21)  & (2.43)  &       &       & (1.33)  & (1.60)  & (2.39)  & (3.03)  & (3.33)  & (2.34) \\
			&       &       &       &       &       &       &       &       &       &       &       &  \\
			ACC2    &   2.16  & 2.15  & 2.89  & 3.62  & 10.14 & 4.19  &       &   ACC2    & 0.67  & 1.76  & 1.31  & 1.93  & 9.71  & 3.08 \\
			&  (1.22)  & (1.61)  & (2.13)  & (2.73)  & (2.27)  & (1.99)  &       &       & (0.42)  & (1.27)  & (1.09)  & (1.68)  & (2.16)  & (1.32) \\
			&       &       &       &       &       &       &       &       &       &       &       &  \\
			ACC3   &      2.46  & 2.49  & 3.28  & 3.32  & 8.65  & 4.04  &       &   ACC3    & 1.68  & 1.86  & 2.79  & 3.28  & 9.42  & 3.81 \\ 
			&  (1.12)  & (2.01)  & (2.77)  & (2.56)  & (2.30)  & (2.15)  &       &       & (0.92)  & (1.62)  & (2.59)  & (2.59)  & (2.51)  & (2.05) \\
			&       &       &       &       &       &       &       &       &       &       &       &  \\
			ACC4    &      3.14  & 0.85  & 2.22  & 2.47  & 7.40  & 3.22  &       &  ACC4     & 1.31  & 0.05  & 2.04  & 1.60  & 8.48  & 2.70 \\
			&  (1.54)  & (0.91)  & (2.31)  & (2.11)  & (2.41)  & (1.86)  &       &       & (0.89)  & (0.06)  & (2.54)  & (1.34)  & (2.75)  & (1.52) \\
			&       &       &       &       &       &       &       &       &       &       &       &  \\
			
			ACC5    &     -0.84 & -0.42 & 0.34  & 0.64  & 3.73  & 0.69  &       &  ACC5     & -1.72 & -0.75 & 0.33  & 0.86  & 5.51  & 0.84 \\ 
			&  (0.82) & (0.62) & (0.57)  & (0.96)  & (1.68)  & (0.35)  &       &       & (1.94) & (1.15) & (0.56)  & (1.30)  & (2.25)  & (0.21) \\
			&       &       &       &       &       &       &       &       &       &       &       &  \\
			
			\textit{top-bottom}    &      6.10  & 5.23  & 4.32  & 5.15  & 13.37 & 6.83  &       &    \textit{top-bottom}   & 6.09  & 3.57  & 4.05  & 5.06  & 12.47 & 6.25 \\
			&  (1.81) & (2.37) & (2.15) & (2.49) & (3.13) & (3.14)  &       &       & (1.84) & (1.99) & (2.27) & (2.53) & (2.85) & (2.99) \\
			&       &       &       &       &       &       &       &       &       &       &       &  \\
			\hline
			\\
			&       &     &  \multicolumn{11}{c}{Panel B: Sort funds by past-$\hat{\alpha}$ and then by Core $ACC$}   \\            
&     &     &     &     &     &   &       &  &     &     &     &    &   &  \\
$\hat{\alpha}_{3f}$ $\rightarrow$ & Q1    & Q2    & Q3    & Q4    & Q5    & Avg  &       & $\hat{\alpha}_{5f}$ $\rightarrow$ & Q1    & Q2    & Q3    & Q4    & Q5    & Avg \\
$\downarrow$ $ACC$&     &     &     &     &     &   &       & $\downarrow$ $ACC$  &     &     &     &     &     &  \\
			\\
			\hline
			\\
			ACC1    & 5.21  & 4.83  & 4.79  & 5.89  & 17.21 & 7.59  &       &    ACC1   & 4.57  & 2.76  & 4.29  & 6.10  & 18.13 & 7.17 \\
			& (1.54)  & (2.30)  & (2.27)  & (2.78)  & (3.22)  & (2.42)  &       &    & (1.40)  & (1.56)  & (2.29)  & (3.06)  & (3.35)  & (2.33) \\
            &       &       &       &       &       &       &       &       &       &       &       &  \\
			ACC2 & 2.09  & 2.11  & 3.09  & 3.64  & 10.32 & 4.25  &       &     ACC2  & 0.90  & 1.66  & 1.24  & 2.03  & 9.38  & 3.04 \\
			& (1.20)  & (1.59)  & (2.29)  & (2.76)  & (2.32)  & (2.03)  &       &       & (0.58)  & (1.21)  & (1.01)  & (1.76)  & (2.10)  & (1.33) \\
            &       &       &       &       &       &       &       &       &       &       &       &  \\
			ACC3 & 2.47  & 2.41  & 3.44  & 3.58  & 8.75  & 4.13  &       &   ACC3    & 1.81  & 1.88  & 2.75  & 3.32  & 8.97  & 3.75 \\
			& (1.12)  & (1.91)  & (2.90)  & (2.81)  & (2.31)  & (2.21)  &       &       & (0.98)  & (1.61)  & (2.58)  & (2.58)  & (2.40)  & (2.03) \\
            &       &       &       &       &       &       &       &       &       &       &       &  \\
			ACC4& 3.26  & 0.63  & 2.19  & 2.52  & 7.38  & 3.20  &       &    ACC4   & 1.35  & 0.04  & 2.02  & 1.44  & 8.27  & 2.62 \\
			& (1.59)  & (0.67)  & (2.26)  & (2.13)  & (2.38)  & (1.81)  &       &       & (0.91)  & (0.05)  & (2.47)  & (1.19)  & (2.70)  & (1.47) \\
            &       &       &       &       &       &       &       &       &       &       &       &  \\
			ACC5 & -0.86 & -0.48 & 0.37  & 0.51  & 3.71  & 0.65  &       &    ACC5   & -1.81 & -0.71 & 0.37  & 0.95  & 5.51  & 0.86 \\
			& (0.86) & (0.72)& (0.61)& (0.78)& (1.68)& (0.30)&       &       & (2.03) & (1.07)& (0.63)& (1.41)& (2.25)& (0.24)\\
            &       &       &       &       &       &       &       &       &       &       &       &  \\
			\textit{top-bottom} & 6.08  & 5.31  & 4.42  & 5.38  & 13.50 & 6.94  &       &   \textit{top-bottom}    & 6.38  & 3.48  & 3.92  & 5.15  & 12.63 & 6.31 \\
			& (1.78) & (2.43) & (2.14) & (2.53) & (3.14) & (2.40) &    &       & (1.93) & (1.93) & (2.15) & (2.53) & (2.87) & (2.28) \\
            &       &       &       &       &       &       &       &       &       &       &       &  \\
			\hline
			\hline
	\end{tabular}}%
	\label{tab:doublesorting_pre}%
\end{table}

\begin{table}[ht!]
	\centering
	\caption{\textbf{Double Sorts Funds by Past Performances and $ACC$ - Crisis.} The table shows the resulting 5x5 portfolios' alphas obtained by sorting funds in quintiles firstly by past-$\hat{\alpha}$ performances and then by $ACC$. Panel B uses the core $ACC$ in which we drop those funds in the tails corresponding to both the top and bottom 5 per cent of the $ACC$ distribution in each 5x5 portfolio. The table reports the OLS estimates of each 5x5 portfolio's alpha (in percentage per year) and the corresponding absolute value of the t-statistics (in parentheses). Subscripts $3f$ and $5f$ stand for the three and five factors models used to compute alpha performances (\cite{fama1993common,fama2015five}). To compute past performance we use nine months of lookback period of daily observations on gross returns, which are determined by adding fund management fees to net returns. We then calculate the return of each quintile portfolio over the next three months using daily returns series and equally weighting funds in each 5x5 portfolio. 5x5 portfolios are redefined each quarter and the corresponding three months returns time series are connected across quarters to form a full sample period for each 5x5 portfolio. \textit{top-bottom} is the portfolio obtained investing long in funds belonging to the best-performer quintile portfolio and short funds in the worst-performer quintile, within the same quintile determined in the first sort. Finally, the portfolio denoted as \textit{Avg} invests equally in each of the five \textit{top-bottom} portfolios. The sample period is September 2007 to June 2010.}
	\scalebox{0.60}{  \begin{tabular}{lccccccp{0.2cm}lcccccc}
			\hline
			\hline
			\\
			&       &     &  \multicolumn{11}{c}{Panel A: Sort funds by past-$\hat{\alpha}$ and then by $ACC$}   \\            
			&     &     &     &     &     &   &       &  &     &     &     &    &   &  \\
			$\hat{\alpha}_{3f}$ $\rightarrow$ & Q1    & Q2    & Q3    & Q4    & Q5    & Avg  &       & $\hat{\alpha}_{5f}$ $\rightarrow$ & Q1    & Q2    & Q3    & Q4    & Q5    & Avg \\
			$\downarrow$ $ACC$ &     &     &     &     &     &   &       & $\downarrow$ $ACC$  &     &     &     &     &     &  \\
			
			\\
			\hline
			\\
			ACC1     & -12.91 & -4.93 & -4.14 & -4.96 & -3.28 & -6.04 &       &    ACC1    & -8.29 & -3.34 & -3.55 & -2.42 & -4.39 & -4.40 \\
			& (1.21) & (0.89) & (1.10) & (1.21) & (0.40) & (0.96) &       &       & (0.81) & (0.56) & (0.78) & (0.63) & (0.64) & (0.68) \\
			&       &       &       &       &       &       &       &       &       &       &       &  \\
			
			ACC2         & -8.82 & -6.18 & -3.67 & -1.60 & -1.04 & -4.26 &       &    ACC2    & -9.25 & -5.70 & -2.96 & -3.77 & -1.29 & -4.59 \\
			& (0.96) & (1.25) & (1.11) & (0.48) & (0.19) & (0.80) &       &       & (1.14) & (1.17) & (0.94) & (1.25) & (0.25) & (0.95) \\
			&       &       &       &       &       &       &       &       &       &       &       &  \\
			
			ACC3          & -7.71 & -5.53 & -4.00 & -5.00 & -10.27 & -6.50 &       &    ACC3   & -11.77 & -4.88 & -4.55 & -5.34 & -4.77 & -6.26 \\
			& (1.18) & (1.39) & (1.19) & (1.27) & (1.67) & (1.34) &       &       & (1.73) & (1.16) & (1.26) & (1.42) & (1.01) & (1.32) \\
			&       &       &       &       &       &       &       &       &       &       &       &  \\
			
			ACC4         & -12.55 & -7.12 & -6.09 & -6.27 & -5.03 & -7.41 &       &    ACC4   & -10.65 & -8.75 & -5.63 & -5.38 & -1.97 & -6.48 \\
			& (2.03) & (1.88) & (1.89) & (1.37) & (0.85) & (1.61) &       &       & (1.77) & (2.20) & (1.58) & (1.26) & (0.40) & (1.44) \\
			&       &       &       &       &       &       &       &       &       &       &       &  \\
			
			ACC5         & -11.56 & -6.64 & -6.90 & -5.80 & -5.64 & -7.31 &       &    ACC5   & -10.34 & -8.27 & -6.25 & -6.66 & -5.47 & -7.40 \\
			& (2.63) & (1.88) & (1.72) & (1.50) & (1.28) & (1.80) &       &       & (2.59) & (2.13) & (1.74) & (1.62) & (1.22) & (1.86) \\
			&       &       &       &       &       &       &       &       &       &       &       &  \\
			
			\textit{top-bottom}       & -1.35 & 1.71  & 2.76  & 0.84  & 2.37  & 1.27  &       &   \textit{top-bottom}    & 2.05  & 4.93  & 2.69  & 4.24  & 1.08  & 3.00 \\
			& (0.13)  & (0.33) & (0.70) & (0.20) & (0.30) & (0.26)  &       &       & (0.21) & (0.88) & (0.63) & (1.00) & (0.15) & (0.60) \\
			
			&       &       &       &       &       &       &       &       &       &       &       &  \\
			\hline
			\\
			&       &     &  \multicolumn{11}{c}{Panel B: Sort funds by past-$\hat{\alpha}$ and then by Core $ACC$}   \\            
&     &     &     &     &     &   &       &  &     &     &     &    &   &  \\
$\hat{\alpha}_{3f}$ $\rightarrow$ & Q1    & Q2    & Q3    & Q4    & Q5    & Avg  &       & $\hat{\alpha}_{5f}$ $\rightarrow$ & Q1    & Q2    & Q3    & Q4    & Q5    & Avg \\
$\downarrow$ $ACC$&     &     &     &     &     &   &       & $\downarrow$ $ACC$  &     &     &     &     &     &  \\

\\
\hline
\\
ACC1     & -13.40 & -4.50 & -4.15 & -5.08 & -3.29 & -6.08 &       &   ACC1    & -8.62 & -2.98 & -3.53 & -2.13 & -4.13 & -4.28 \\
& (1.23) & (0.80) & (1.08) & (1.22) & (0.40) & (0.95) &       &       & (0.84) & (0.49) & (0.76) & (0.54) & (0.59) & (0.64) \\
&       &       &       &       &       &       &       &       &       &       &       &  \\
ACC2 & -8.44 & -6.45 & -4.12 & -1.55 & -1.25 & -4.36 &       &   ACC2    & -9.25 & -5.76 & -2.76 & -3.90 & -1.37 & -4.61 \\
& (0.91) & (1.30) & (1.23) & (0.47) & (0.22) & (0.83) &       &       & (1.15) & (1.19) & (0.89) & (1.31) & (0.27) & (0.96) \\
&       &       &       &       &       &       &       &       &       &       &       &  \\
ACC3 & -7.71 & -5.77 & -3.97 & -4.96 & -10.48 & -6.58 &       &   ACC3    & -11.56 & -4.81 & -4.21 & -5.34 & -4.78 & -6.14 \\
& (1.17) & (1.43) & (1.18) & (1.26) & (1.69) & (1.35) &       &       & (1.71) & (1.14) & (1.17) & (1.42) & (1.01) & (1.29) \\
&       &       &       &       &       &       &       &       &       &       &       &  \\
ACC4 & -12.38 & -6.87 & -6.29 & -6.28 & -4.95 & -7.35 &       &    ACC4   & -10.16 & -8.43 & -5.44 & -5.49 & -1.91 & -6.28 \\
& (2.03) & (1.82) & (1.91) & (1.38) & (0.84) & (1.60) &       &       & (1.68) & (2.15) & (1.51) & (1.27) & (0.39) & (1.40) \\
&       &       &       &       &       &       &       &       &       &       &       &  \\
ACC5 & -11.73 & -6.74 & -6.86 & -5.41 & -5.82 & -7.31 &       &     ACC5  & -10.59 & -8.23 & -6.19 & -6.52 & -5.53 & -7.41 \\
& (2.65) & (1.89) & (1.70) & (1.41) & (1.31) & (1.79) &       &       & (2.59) & (2.08) & (1.69) & (1.60) & (1.24) & (1.84) \\
&       &       &       &       &       &       &       &       &       &       &       &  \\
\textit{top-bottom} & -1.67 & 2.24  & 2.71  & 0.33  & 2.53  & 1.23  &       &    \textit{top-bottom}   & 1.97  & 5.25  & 2.66  & 4.39  & 1.40  & 3.14 \\
& (0.15)  & (0.42) & (0.66) & (0.08) & (0.31) & (0.26) &       &       & (0.20) & (0.90) & (0.61) & (1.00) & (0.19) & (0.58) \\
			&       &       &       &       &       &       &       &       &       &       &       &  \\
			\hline
			\hline
	\end{tabular}}%
	\label{tab:doublesorting_post}%
\end{table}

\begin{table}[ht!]
	\centering
	\caption{\textbf{Double Sorts Funds by Past Performances and $ACC$.} The table shows the resulting 5x5 $ACC$portfolios' alphas obtained by sorting funds in quintiles firstly by past-$\hat{\alpha}$ performances and then by$ACC$. Panel B uses the core $ACC$ in which we drop those funds in the tails corresponding to both the top and bottom 5 per cent of the $ACC$ distribution in each 5x5 portfolio. The table reports the OLS estimates of each 5x5 portfolio's alpha (in percentage per year) and the corresponding absolute value of the t-statistics (in parentheses). Subscripts $3f$ and $5f$ stand for the three and five factors models used to compute alpha performances (\cite{fama1993common,fama2015five}). To compute past performance we use nine months of lookback period of daily observations on gross returns, which are determined by adding fund management fees to net returns. We then calculate the return of each quintile portfolio over the next three months using daily returns series and equally weighting funds in each 5x5 portfolio. 5x5 portfolios are redefined each quarter and the corresponding three months returns time series are connected across quarters to form a full sample period for each 5x5 portfolio. \textit{top-bottom} is the portfolio obtained investing long in funds belonging to the best-performer quintile portfolio and short funds in the worst-performer quintile, within the same quintile determined in the first sort. Finally, the portfolio denoted as \textit{Avg} invests equally in each of the five \textit{top-bottom} portfolios. The sample period is June 2004 to June 2010.}
	\scalebox{0.60}{  \begin{tabular}{lccccccp{0.2cm}lcccccc}
			\hline
			\hline
			\\
			&       &     &  \multicolumn{11}{c}{Panel A: Sort funds by past-$\hat{\alpha}$ and then by $ACC$}   \\            
			&     &     &     &     &     &   &       &  &     &     &     &    &   &  \\
			$\hat{\alpha}_{3f}$ $\rightarrow$ & Q1    & Q2    & Q3    & Q4    & Q5    & Avg  &       & $\hat{\alpha}_{5f}$ $\rightarrow$ & Q1    & Q2    & Q3    & Q4    & Q5    & Avg \\
			$\downarrow$ $ACC$&     &     &     &     &     &   &       & $\downarrow$ $ACC$  &     &     &     &     &     &  \\
			
			\\
			\hline
			\\
			ACC1    & -0.89 & 1.41  & 0.97  & 0.78  & 6.85  & 1.82  &       &      ACC1    & 0.43  & 0.91  & 1.33  & 2.06  & 7.06  & 2.36 \\
			& (0.16) & (0.49)  & (0.47)  & (0.35)  & (1.41)  & (0.51)  &       &       & (0.08)  & (0.30)  & (0.55)  & (0.99)  & (1.59)  & (0.70) \\
			&       &       &       &       &       &       &       &       &       &       &       &  \\
			
		ACC2              & -2.30 & -1.45 & 0.28  & 1.49  & 4.87  & 0.58  &       &      ACC2   & -3.27 & -1.68 & -0.51 & -0.48 & 4.53  & -0.28 \\
			& (0.51) & (0.60) & (0.16)  & (0.88)  & (1.38)  & (0.26)  &       &       & (0.83) & (0.71) & (0.32) & (0.31) & (1.32)  & (0.17) \\
			&       &       &       &       &       &       &       &       &       &       &       &  \\
			
			ACC3              & -2.11 & -1.27 & 0.04  & -0.15 & 0.09  & -0.68 &       &      ACC3   & -4.25 & -1.16 & -0.55 & -0.39 & 3.04  & -0.66 \\
			& (0.65) & (0.65) & (0.03)  & (0.07) & (0.03)  & (0.27) &       &       & (1.30) & (0.57) & (0.32) & (0.21) & (1.02)  & (0.28) \\
			&       &       &       &       &       &       &       &       &       &       &       &  \\
			
			ACC4              & -4.61 & -2.74 & -1.43 & -0.94 & 2.15  & -1.51 &       &    ACC4     & -4.19 & -3.78 & -1.21 & -1.22 & 4.19  & -1.24 \\
			& (1.50) & (1.48) & (0.91) & (0.43) & (0.67)  & (0.73) &       &       & (1.46) & (2.00) & (0.71) & (0.59) & (1.49)  & (0.65) \\
			&       &       &       &       &       &       &       &       &       &       &       &  \\
			
			ACC5              & -6.61 & -3.32 & -2.84 & -1.84 & -0.02 & -2.93 &       &      ACC5   & -5.87 & -3.96 & -2.35 & -2.10 & 1.37  & -2.58 \\
			& (3.04) & (1.94) & (1.49) & (1.00) & (0.01) & (1.49) &       &       & (3.05) & (2.15) & (1.37) & (1.08) & (0.55)  & (1.42) \\
			&       &       &       &       &       &       &       &       &       &       &       &  \\
			
			\textit{top-bottom}        & 5.72  & 4.73  & 3.81  & 2.62  & 6.87  & 4.75  &       &  \textit{top-bottom}    & 6.30  & 4.87  & 3.67  & 4.15  & 5.69  & 4.94 \\
			& (1.04) & (1.68) & (1.74) & (1.13) & (1.52) & (1.87)  &       &       & (1.22) & (1.67) & (1.59) & (1.81) & (1.32) & (1.89) \\
			&       &       &       &       &       &       &       &       &       &       &       &  \\
			\hline
			\\
			&       &     &  \multicolumn{11}{c}{Panel B: Sort funds by past-$\hat{\alpha}$ and then by Core $ACC$}   \\            
&     &     &     &     &     &   &       &  &     &     &     &    &   &  \\
$\hat{\alpha}_{3f}$ $\rightarrow$ & Q1    & Q2    & Q3    & Q4    & Q5    & Avg  &       & $\hat{\alpha}_{5f}$ $\rightarrow$ & Q1    & Q2    & Q3    & Q4    & Q5    & Avg \\
$\downarrow$ $ACC$&     &     &     &     &     &   &       & $\downarrow$ $ACC$  &     &     &     &     &     &  \\
\\
			\hline
			\\
			 ACC1    & -1.14 & 1.59  & 1.05  & 0.77  & 6.91  & 1.84  &       &    ACC1   & 0.39  & 1.08  & 1.32  & 2.29  & 7.23  & 2.46 \\
			& (0.21) & (0.55)  & (0.49)  & (0.34)  & (1.41)  & (0.52)  &       &       & (0.07)  & (0.35)  & (0.54)  & (1.08)  & (1.61)  & (0.73) \\
            &       &       &       &       &       &       &       &       &       &       &       &  \\
			ACC2 & -2.17 & -1.60 & 0.16  & 1.52  & 4.92  & 0.57  &       &    ACC2   & -3.15 & -1.81 & -0.47 & -0.48 & 4.34  & -0.31 \\
			& (0.48) & (0.66) & (0.09) & (0.90)  & (1.39)  & (0.25)  &       &       & (0.81) & (0.77) & (0.29) & (0.32) & (1.28)  & (0.18) \\
            &       &       &       &       &       &       &       &       &       &       &       &  \\
			ACC3 & -2.07 & -1.43 & 0.15  & -0.01 & 0.07  & -0.66 &       &     ACC3  & -4.07 & -1.12 & -0.44 & -0.38 & 2.80  & -0.64 \\
			& (0.63) & (0.73) & (0.09) & (0.00)  & (0.02)  & (0.25) &       &       & (1.25) & (0.55) & (0.25) & (0.20) & (0.94)  & (0.26) \\
            &       &       &       &       &       &       &       &       &       &       &       &  \\
		ACC4 	& -4.41 & -2.73 & -1.51 & -0.90 & 2.17  & -1.48 &       &     ACC4  & -3.96 & -3.61 & -1.13 & -1.36 & 4.10  & -1.19 \\
			& (1.45) & (1.47) & (0.94)& (0.41) & (0.68)  & (0.72) &       &       & (1.38) & (1.94) & (0.66) & (0.65) & (1.47)  & (0.63) \\
            &       &       &       &       &       &       &       &       &       &       &       &  \\
		ACC5 	& -6.69 & -3.38 & -2.79 & -1.73 & -0.09 & -2.94 &       &     ACC5  & -6.05 & -3.92 & -2.31 & -1.98 & 1.35  & -2.58 \\
			& (3.06) & (1.96) & (1.46)& (0.95) & (0.04) & (1.49) &       &       & (3.08) & (2.09) & (1.32) & (1.03) & (0.54)  & (1.39) \\
            &       &       &       &       &       &       &       &       &       &       &       &  \\
		\textit{top-bottom}	& 5.55  & 4.98  & 3.84  & 2.51  & 7.00  & 4.77  &       &   \textit{top-bottom}    & 6.44  & 5.00  & 3.63  & 4.27  & 5.88  & 5.04 \\
			& (1.00) & (1.74) & (1.70) & (1.06) & (1.53) & (1.41) &       &       & (1.24) & (1.67) & (1.54) & (1.81) & (1.35) & (1.52) \\
			
			&       &       &       &       &       &       &       &       &       &       &       &  \\
			\hline
			\hline
	\end{tabular}}%
	\label{tab:double_all}%
\end{table}

\clearpage
\subsection{Reverse Sorts}
\label{rs}

Findings from Tables \ref{tab:doublesorting_pre}-\ref{tab:double_all} indicate that on average higher values of specialization (low $ACC$ value) produce better extra-performances, although this relationship is influenced by both the quintiles chosen in the first sort and the impact of the crisis of mid-2007. For this reason we also analyze whether information about funds' future performances is contained in past-$\hat{\alpha}$ and not in the $ACC$ distribution. Hence, we sort funds in quintile in reverse, i.e. firstly by the $ACC$ and then by past-$\hat{\alpha}$, and we report in Tables \ref{tab:reverse_pre}-\ref{tab:reverse_all} the resulting 5x5 portfolios as well as the \textit{Avg} and \textit{top-bottom} strategies for different time windows. 

The average difference between \textit{top-bottom} portfolios prior to the crisis is very significant and above 7 per cent in both the three and five factors models (see Table \ref{tab:reverse_pre}). This means that past-$\hat{\alpha}$ sort adds incremental information about future funds performances beyond the $ACC$ index especially for those less common portfolios (namely, those in ACC1). Hence, prior to the crisis of mid-2007, double-sorting funds according to the $ACC$ index and then past-$\hat{\alpha}$ produces 5x5 quintile portfolios' extra-performances on average comparable with those obtained applying the opposite double-sorts criterion. By contrast, for observations in the interval 2007-2010 we note in Table \ref{tab:reverse_post} that sorting firstly by the $ACC$ index and then by past-$\hat{\alpha}$ determines higher performances than the opposite case (see Table \ref{tab:doublesorting_post}). During the outbreak of financial markets, therefore, persistence in past-$\hat{\alpha}$ seems to have driven better quintile extra-performances than the $ACC$ property. These results are supported also by the enlarged time window from 2004 to 2010 (see Table \ref{tab:reverse_all}). Once again by combing information from both past-$\hat{\alpha}$ and $ACC$, investors can benefit from the investment strategy long in the \textit{top} double-sort quintile and short in the \textit{bottom} one, thus supporting the use of the proposed topological indicator as a complementary information that can be exploited to build portfolios. These results are largely confirmed by the \textit{Core} cases (see panels B in Tables \ref{tab:reverse_pre}-\ref{tab:reverse_all}).

\begin{table}[ht!]
	\centering
	\caption{\textbf{Reverse Double Sorts Funds by $ACC$ and Past Performances - Pre Crisis.} The table shows in Panel A the resulting 5x5 portfolios' alphas obtained by sorting funds in quintiles firstly by $ACC$ and then by past-$\hat{\alpha}$ performances. Panel B uses the core $ACC$ in which we drop those funds in the tails corresponding to both the top and bottom 5 per cent of the $ACC$ distribution in each 5x5 portfolio. The table reports the OLS estimates of each 5x5 portfolio's alpha (in percentage per year) and the corresponding absolute value of the t-statistics (in parentheses). Subscripts $3f$ and $5f$ stand for the three and five factors models used to compute alpha performances (\cite{fama1993common,fama2015five}). To compute past performance we use nine months of lookback period of daily observations on gross returns, which are determined by adding fund management fees to net returns. We then calculate the return of each quintile portfolio over the next three months using daily returns series and equally weighting funds in each 5x5 portfolio. 5x5 portfolios are redefined each quarter and the corresponding three months returns time series are connected across quarters to form a full sample period for each 5x5 portfolio. \textit{top-bottom} is the portfolio obtained investing long in funds belonging to the best-performer quintile portfolio and short funds in the worst-performer quintile, within the same quintile determined in the first sort. Finally, the portfolio denoted as \textit{Avg} invests equally in each of the five \textit{top-bottom} portfolios. The sample period is June 2004 to June 2007.}
	\scalebox{0.60}{  \begin{tabular}{lccccccp{0.2cm}lcccccc}
			\hline
			\hline
			\\
			&       &     &  \multicolumn{11}{c}{Panel A: Sorting funds by $ACC$ and then by past-$\hat{\alpha}$} \\              
			&     &     &     &     &     &   &       &  &     &     &     &    &   &  \\
			$ACC$  $\rightarrow$ & ACC1    & ACC2    & ACC3    & ACC4    & ACC5    & Avg  &       & $ACC$  $\rightarrow$ & ACC1    & ACC2    & ACC3    & ACC4    & ACC5    & Avg \\
			$\downarrow$ $\hat{\alpha}_{3f}$&     &     &     &     &     &   &       & $\downarrow$ $\hat{\alpha}_{5f}$  &     &     &     &     &     &  \\
			
			\\
			\hline
			\\
			Q1    & 3.88  & 3.44  & 1.74  & -0.16 & -0.93 & 1.59  &       &   Q1    & 3.25  & 2.70  & 1.31  & -0.53 & -1.04 & 1.14 \\
			& (1.77)  & (1.51)  & (1.44)  & (0.11) & (0.93) & (0.74)  &       &       & (1.51)  & (1.22)  & (1.12)  & (0.36) & (1.08) & (0.48) \\
			&       &       &       &       &       &       &       &       &       &       &       &  \\
			
			Q2       & 6.25  & 0.94  & 0.81  & 0.54  & -0.64 & 1.58  &       &  Q2     & 6.01  & 0.56  & 0.70  & 0.30  & -0.73 & 1.37 \\
			& (2.42)  & (0.89)  & (1.03)  & (0.63)  & (0.84) & (0.83)  &       &       & (2.33)  & (0.55)  & (0.89)  & (0.36)  & (0.99) & (0.63) \\
			&       &       &       &       &       &       &       &       &       &       &       &  \\
			
			Q3        & 7.29  & 3.08  & 2.04  & 1.73  & -0.11 & 2.81  &       &    Q3   & 7.11  & 2.84  & 1.79  & 1.52  & -0.14 & 2.62 \\
			& (2.11)  & (3.47)  & (2.46)  & (2.12)  & (0.15) & (2.00)  &       &       & (2.05)  & (3.28)  & (2.20)  & (1.92)  & (0.20) & (1.85) \\
			&       &       &       &       &       &       &       &       &       &       &       &  \\
			
			Q4       & 8.86  & 3.72  & 1.07  & 2.17  & 0.98  & 3.36  &       &   Q4    & 8.69  & 3.48  & 0.94  & 1.93  & 0.87  & 3.18 \\
			& (2.23)  & (3.12)  & (1.33)  & (2.58)  & (1.27)  & (2.11)  &       &       & (2.18)  & (2.94)  & (1.19)  & (2.35)  & (1.16)  & (1.97) \\
			&       &       &       &       &       &       &       &       &       &       &       &  \\
			
			Q5       & 16.73 & 9.66  & 6.87  & 5.91  & 5.44  & 8.92  &       &   Q5    & 16.20 & 9.15  & 6.47  & 5.55  & 5.11  & 8.50 \\
			& (3.17)  & (3.05)  & (2.99)  & (2.87)  & (2.61)  & (2.94)  &       &       & (3.07)  & (2.92)  & (2.87)  & (2.75)  & (2.48)  & (2.82) \\
			&       &       &       &       &       &       &       &       &       &       &       &  \\
			
			\textit{top-bottom}         & 12.86 & 6.21 & 5.13 & 6.08 & 6.37 & 7.33  &       &     \textit{top-bottom}   & 12.95 & 6.44 & 5.16 & 6.08 & 6.15 & 7.36 \\
			& (2.82)  & (1.72)  & (2.25)  & (2.58)  & (2.65)  & (2.79) &       &       & (2.84)  & (1.80)  & (2.29)  & (2.59)  & (2.57)  & (2.81) \\
			&       &       &       &       &       &       &       &       &       &       &       &  \\
			\hline
			\\
			&       &     &  \multicolumn{11}{c}{Panel B: Sorting funds by Core $ACC$ and then by past-$\hat{\alpha}$} \\              
&     &     &     &     &     &   &       &  &     &     &     &    &   &  \\
$ACC$ $\rightarrow$ & ACC1    & ACC2    & ACC3    & ACC4    & ACC5    & Avg  &       & $ACC$ $\rightarrow$ & ACC1    & ACC2    & ACC3    & ACC4    & ACC5    & Avg \\
$\downarrow$ $\hat{\alpha}_{3f}$&     &     &     &     &     &   &       & $\downarrow$ $\hat{\alpha}_{5f}$  &     &     &     &     &     &  \\
			\\
			\hline
			\\
			Q1    & -0.04 & 3.07  & 2.22  & -0.04 & -1.08 & 0.83  &       &   Q1    & -0.42 & 2.35  & 1.80  & -0.42 & -1.15 & 0.43 \\
          & (0.03) & (1.33)  & (1.75)  & (0.03) & (1.00) & (0.41)  &       &       & (0.28) & (1.05)  & (1.46)  & (0.28) & (1.10) & (0.17) \\
          	&       &       &       &       &       &       &       &       &       &       &       &  \\
         Q2 & 0.15  & 1.17  & 0.55  & 0.15  & -0.52 & 0.30  &       &    Q2   & -0.09 & 0.83  & 0.44  & -0.09 & -0.65 & 0.09 \\
          & (0.17)  & (1.09)  & (0.65)  & (0.17)  & (0.62) & (0.29)  &       &       & (0.10) & (0.79)  & (0.53)  & (0.10) & (0.80) & (0.06) \\
          	&       &       &       &       &       &       &       &       &       &       &       &  \\
         Q3 & 1.64  & 3.55  & 1.75  & 1.64  & -0.40 & 1.63  &       &    Q3   & 1.44  & 3.32  & 1.50  & 1.44  & -0.42 & 1.46 \\
          & (1.95)  & (3.79)  & (2.16)  & (1.95)  & (0.55) & (1.86)  &       &       & (1.78)  & (3.63)  & (1.89)  & (1.78)  & (0.58) & (1.70) \\
          	&       &       &       &       &       &       &       &       &       &       &       &  \\
        Q4  & 2.33  & 3.51  & 1.23  & 2.33  & 0.58  & 1.99  &       &     Q4  & 2.02  & 3.30  & 1.07  & 2.02  & 0.50  & 1.78 \\
          & (2.59)  & (3.09)  & (1.45)  & (2.59)  & (0.74)  & (2.09)  &       &       & (2.33)  & (2.93)  & (1.28)  & (2.33)  & (0.66)  & (1.90) \\
          	&       &       &       &       &       &       &       &       &       &       &       &  \\
        Q5  & 5.77  & 10.38 & 7.38  & 5.77  & 4.69  & 6.80  &       &     Q5  & 5.46  & 9.83  & 6.76  & 5.46  & 4.44  & 6.39 \\
          & (2.71)  & (2.83)  & (2.28)  & (2.71)  & (2.46)  & (2.60)  &       &       & (2.61)  & (2.70)  & (2.11)  & (2.61)  & (2.36)  & (2.48) \\
          	&       &       &       &       &       &       &       &       &       &       &       &  \\
       \textit{top-bottom}    & 5.81 & 7.32 & 5.16 & 5.81 & 5.77 & 5.97 &       &     \textit{top-bottom}   & 5.88 & 7.48 & 4.95 & 5.88 & 5.59 & 5.96 \\
          & (2.37)  & (1.81)  & (1.64)  & (2.37)  & (2.56)  & (2.15)  &       &       & (2.40)  & (1.86)  & (1.57)  & (2.40)  & (2.49)  & (2.14) \\
						&       &       &       &       &       &       &       &       &       &       &       &  \\
			\hline
			\hline
	\end{tabular}}%
	\label{tab:reverse_pre}%
\end{table}

\begin{table}[ht!]
	\centering 
	\caption{\textbf{Reverse Double Sorts Funds by $ACC$ and Past Performances - Crisis.} The table shows in Panel A the resulting 5x5 portfolios' alphas obtained by sorting funds in quintiles firstly by $ACC$ and then by past-$\hat{\alpha}$ performances. Panel B uses the core $ACC$ in which we drop those funds in the tails corresponding to both the top and bottom 5 per cent of the $ACC$ distribution in each 5x5 portfolio. The table reports the OLS estimates of each 5x5 portfolio's alpha (in percentage per year) and the corresponding absolute value of the t-statistics (in parentheses). Subscripts $3f$ and $5f$ stand for the three and five factors models used to compute alpha performances (\cite{fama1993common,fama2015five}). To compute past performance we use nine months of lookback period of daily observations on gross returns, which are determined by adding fund management fees to net returns. We then calculate the return of each quintile portfolio over the next three months using daily returns series and equally weighting funds in each 5x5 portfolio. 5x5 portfolios are redefined each quarter and the corresponding three months returns time series are connected across quarters to form a full sample period for each 5x5 portfolio. \textit{top-bottom} is the portfolio obtained investing long in funds belonging to the best-performer quintile portfolio and short funds in the worst-performer quintile, within the same quintile determined in the first sort. Finally, the portfolio denoted as \textit{Avg} invests equally in each of the five \textit{top-bottom} portfolios. The sample period is September 2007 to June 2010.}
	\scalebox{0.60}{  \begin{tabular}{lccccccp{0.2cm}lcccccc}
			\hline
			\hline
			\\
			&       &     &  \multicolumn{11}{c}{Panel A: Sorting funds by $ACC$ and then by past-$\hat{\alpha}$} \\              
			&     &     &     &     &     &   &       &  &     &     &     &    &   &  \\
			$ACC$ $\rightarrow$ & ACC1    & ACC2    & ACC3    & ACC4    & ACC5    & Avg  &       & $ACC$ $\rightarrow$ & ACC1    & ACC2    & ACC3    & ACC4    & ACC5    & Avg \\
			$\downarrow$ $\hat{\alpha}_{3f}$&     &     &     &     &     &   &       & $\downarrow$ $\hat{\alpha}_{5f}$  &     &     &     &     &     &  \\
			
			\\
			\hline
			\\
			Q1    & -13.38 & -13.89 & -13.57 & -13.79 & -10.98 & -13.12 &       &    Q1   & -10.78 & -12.56 & -12.17 & -12.42 & -9.16 & -11.42 \\
			& (1.85) & (2.20) & (2.63) & (2.82) & (1.50) & (2.20) &       &       & (1.50) & (2.00) & (2.39) & (2.63) & (1.25) & (1.95) \\
			&       &       &       &       &       &       &       &       &       &       &       &  \\
			
			Q2          & -7.36 & -4.01 & -7.91 & -7.75 & -7.29 & -6.86 &       &    Q2   & -6.52 & -3.70 & -7.43 & -7.15 & -7.02 & -6.37 \\
			& (1.07) & (1.09) & (2.07) & (2.05) & (2.10) & (1.68) &       &       & (0.95) & (1.02) & (1.97) & (1.91) & (2.05) & (1.58) \\
			&       &       &       &       &       &       &       &       &       &       &       &  \\
			
			Q3         & -8.46 & -4.01 & -5.34 & -7.80 & -4.62 & -6.05 &       &   Q3    & -7.39 & -4.00 & -5.21 & -7.38 & -4.79 & -5.75 \\
			& (1.19) & (1.19) & (1.64) & (2.13) & (1.37) & (1.51) &       &       & (1.04) & (1.19) & (1.63) & (2.04) & (1.44) & (1.47) \\
			&       &       &       &       &       &       &       &       &       &       &       &  \\
			
			Q4          & -4.17 & -3.99 & -5.80 & -5.74 & -5.37 & -5.01 &       &   Q4    & -3.75 & -3.81 & -5.49 & -5.83 & -5.52 & -4.88 \\
			& (0.77) & (1.31) & (1.63) & (1.67) & (1.33) & (1.34) &       &       & (0.70) & (1.28) & (1.56) & (1.74) & (1.37) & (1.33) \\
			&       &       &       &       &       &       &       &       &       &       &       &  \\
			
			Q5         & -4.11 & -1.84 & -3.02 & -1.42 & -4.92 & -3.06 &       &    Q5   & -3.58 & -2.61 & -3.47 & -2.43 & -6.21 & -3.66 \\
			& (0.66) & (0.39) & (0.68) & (0.31) & (1.21) & (0.65) &       &       & (0.58) & (0.57) & (0.81) & (0.58) & (1.66) & (0.84) \\
			&       &       &       &       &       &       &       &       &       &       &       &  \\
			
			\textit{top-bottom}         & 9.27 & 12.06 & 10.55 & 12.38 & 6.06 & 10.06 &       &    \textit{top-bottom}   & 7.20 & 9.95 & 8.70 & 9.99 & 2.95 & 7.76 \\
			& (1.40)  & (1.90)  & (2.16)  & (2.58)  & (0.71)  & (1.75) &       &       & (1.09)  & (1.57)  & (1.80)  & (2.11)  & (0.35)  & (1.52) \\
			&       &       &       &       &       &       &       &       &       &       &       &  \\
			\hline
			\\
			&       &     &  \multicolumn{11}{c}{Panel B: Sorting funds by Core $ACC$ and then by past-$\hat{\alpha}$} \\              
&     &     &     &     &     &   &       &  &     &     &     &    &   &  \\
$ACC$ $\rightarrow$ & ACC1    & ACC2    & ACC3    & ACC4    & ACC5    & Avg  &       & $ACC$ $\rightarrow$ & ACC1    & ACC2    & ACC3    & ACC4    & ACC5    & Avg \\
$\downarrow$ $\hat{\alpha}_{3f}$&     &     &     &     &     &   &       & $\downarrow$ $\hat{\alpha}_{5f}$  &     &     &     &     &     &  \\
			\\
			\hline
			\\
			Q1    & -13.35 & -12.52 & -13.27 & -13.35 & -11.15 & -12.73 &       &    Q1   & -12.10 & -11.24 & -11.92 & -12.10 & -9.28 & -11.33 \\
          & (2.72) & (2.05) & (2.62) & (2.72) & (1.54) & (2.33) &       &       & (2.55) & (1.86) & (2.38) & (2.55) & (1.28) & (2.12) \\
          &       &       &       &       &       &       &       &       &       &       &       &  \\
Q2          & -7.69 & -3.77 & -7.83 & -7.69 & -7.76 & -6.95 &       &    Q2   & -7.02 & -3.46 & -7.43 & -7.02 & -7.47 & -6.48 \\
          & (1.98) & (1.08) & (1.97) & (1.98) & (2.24) & (1.85) &       &       & (1.83) & (1.00) & (1.89) & (1.83) & (2.18) & (1.75) \\
          &       &       &       &       &       &       &       &       &       &       &       &  \\
Q3          & -7.38 & -5.23 & -5.94 & -7.38 & -5.28 & -6.24 &       &     Q3  & -6.86 & -5.27 & -5.76 & -6.86 & -5.69 & -6.08 \\
          & (2.15) & (1.53) & (1.74) & (2.15) & (1.50) & (1.81) &       &       & (2.03) & (1.55) & (1.71) & (2.03) & (1.64) & (1.79) \\
          &       &       &       &       &       &       &       &       &       &       &       &  \\
 Q4         & -5.45 & -2.41 & -4.41 & -5.45 & -5.32 & -4.61 &       &   Q4    & -5.65 & -2.21 & -4.31 & -5.65 & -5.35 & -4.63 \\
          & (1.56) & (0.90) & (1.32) & (1.56) & (1.36) & (1.34) &       &       & (1.65) & (0.84) & (1.30) & (1.65) & (1.38) & (1.36) \\
          &       &       &       &       &       &       &       &       &       &       &       &  \\
  Q5        & -1.22 & -1.28 & -3.62 & -1.22 & -5.93 & -2.65 &       &    Q5   & -2.21 & -2.31 & -3.97 & -2.21 & -7.12 & -3.57 \\
          & (0.27) & (0.27) & (0.78) & (0.27) & (1.55) & (0.63) &       &       & (0.53) & (0.49) & (0.90) & (0.53) & (2.01) & (0.89) \\
          &       &       &       &       &       &       &       &       &       &       &       &  \\
  \textit{top-bottom}         & 12.13 & 11.24 & 9.65 & 12.13 & 5.22 & 10.07 &       &    \textit{top-bottom}    & 9.88 & 8.94 & 7.96 & 9.88 & 2.16 & 7.76 \\
          & (2.46)  & (1.70)  & (2.03)  & (2.46)  & (0.63)  & (1.86)  &       &       & (2.02)  & (1.35)  & (1.69)  & (2.02)  & (0.27)  & (1.47) \\
          &       &       &       &       &       &       &       &       &       &       &       &  \\

			\hline
			\hline
	\end{tabular}}%
	\label{tab:reverse_post}%
\end{table}

\begin{table}[ht!]
	\centering
	\caption{\textbf{Reverse Double Sorts Funds by $ACC$ and Past Performances.} The table shows in Panel A the resulting 5x5 portfolios' alphas obtained by sorting funds in quintiles firstly by $ACC$ and then by past-$\hat{\alpha}$ performances. Panel B uses the core $ACC$ in which we drop those funds in the tails corresponding to both the top and bottom 5 per cent of the $ACC$ distribution in each 5x5 portfolio. The table reports the OLS estimates of each 5x5 portfolio's alpha (in percentage per year) and the corresponding absolute value of the t-statistics (in parentheses). Subscripts $3f$ and $5f$ stand for the three and five factors models used to compute alpha performances (\cite{fama1993common,fama2015five}). To compute past performance we use nine months of lookback period of daily observations on gross returns, which are determined by adding fund management fees to net returns. We then calculate the return of each quintile portfolio over the next three months using daily returns series and equally weighting funds in each 5x5 portfolio. 5x5 portfolios are redefined each quarter and the corresponding three months returns time series are connected across quarters to form a full sample period for each 5x5 portfolio. \textit{top-bottom} is the portfolio obtained investing long in funds belonging to the best-performer quintile portfolio and short funds in the worst-performer quintile, within the same quintile determined in the first sort. Finally, the portfolio denoted as \textit{Avg} invests equally in each of the five \textit{top-bottom} portfolios. The sample period is June 2004 to June 2010.}
	\scalebox{0.60}{  \begin{tabular}{lccccccp{0.2cm}lcccccc}
			\hline
			\hline
			\\
			&       &     &  \multicolumn{11}{c}{Panel A: Sorting funds by $ACC$ and then by past-$\hat{\alpha}$} \\              
			&     &     &     &     &     &   &       &  &     &     &     &    &   &  \\
			$ACC$ $\rightarrow$ & ACC1    & ACC2    & ACC3    & ACC4    & ACC5    & Avg  &       & $ACC$ $\rightarrow$ & ACC1    & ACC2    & ACC3    & ACC4    & ACC5    & Avg \\
			$\downarrow$ $\hat{\alpha}_{3f}$&     &     &     &     &     &   &       & $\downarrow$ $\hat{\alpha}_{5f}$  &     &     &     &     &     &  \\
			
			\\
			\hline
			\\
			Q1    & -2.23 & -4.05 & -6.41 & -4.33 & -4.84 & -4.37 &       &   Q1    & -1.21 & -3.68 & -6.26 & -4.37 & -4.66 & -4.04 \\
			& (0.48) & (1.28) & (2.34) & (1.75) & (2.72) & (1.71) &       &       & (0.26) & (1.16) & (2.32) & (1.80) & (2.63) & (1.64) \\
			&       &       &       &       &       &       &       &       &       &       &       &  \\
			
			Q2          & -0.06 & -2.76 & -1.99 & -1.47 & -3.92 & -2.04 &       &    Q2   & 0.34  & -2.68 & -1.79 & -1.54 & -3.97 & -1.93 \\
			& (0.02) & (1.28) & (0.89) & (0.84) & (2.07) & (1.02) &       &       & (0.10)  & (1.27) & (0.82) & (0.90) & (2.09) & (1.00) \\
			&       &       &       &       &       &       &       &       &       &       &       &  \\
			
			Q3          & 1.30  & -0.40 & 0.93  & -1.03 & -3.66 & -0.57 &       &    Q3   & 1.96  & -0.53 & 0.74  & -1.11 & -3.80 & -0.55 \\
			& (0.35)  & (0.27) & (0.54)  & (0.58) & (1.91) & (0.37) &       &       & (0.54)  & (0.39) & (0.45)  & (0.63) & (1.99) & (0.40) \\
			&       &       &       &       &       &       &       &       &       &       &       &  \\
			
			Q4          & 4.27  & 0.18  & 0.62  & -0.14 & -2.56 & 0.47  &       &   Q4    & 4.51  & 0.00  & 0.35  & -0.02 & -2.62 & 0.44 \\
			& (1.29)  & (0.13)  & (0.35)  & (0.08) & (1.33) & (0.07)  &       &       & (1.37)  & (0.00)  & (0.20)  & (0.01) & (1.36) & (0.04) \\
			&       &       &       &       &       &       &       &       &       &       &       &  \\
			
			Q5          & 7.86  & 5.56  & 3.98  & 2.84  & -1.34 & 3.78  &       &   Q5    & 6.99  & 5.55  & 3.79  & 3.40  & -1.05 & 3.74 \\
			& (1.67)  & (2.08)  & (1.56)  & (1.16)  & (0.65) & (1.16)  &       &       & (1.52)  & (2.12)  & (1.52)  & (1.42)  & (0.51) & (1.22) \\
			&       &       &       &       &       &       &       &       &       &       &       &  \\
			
			\textit{top-bottom}         & 10.09 & 9.61 & 10.40 & 7.17 & 3.49 & 8.15 &       &   \textit{top-bottom}    & 8.20 & 9.23 & 10.05 & 7.77 & 3.61 & 7.77 \\
			& (1.94)  & (2.45)  & (3.40)  & (2.62)  & (2.60)  & (3.18) &       &       & (1.59)  & (2.36)  & (3.28)  & (2.84)  & (2.69)  & (3.03) \\
			&       &       &       &       &       &       &       &       &       &       &       &  \\
			\hline
			\\
			&       &     &  \multicolumn{11}{c}{Panel B: Sorting funds by Core $ACC$ and then by past-$\hat{\alpha}$} \\              
&     &     &     &     &     &   &       &  &     &     &     &    &   &  \\
$ACC$ $\rightarrow$ & ACC1    & ACC2    & ACC3    & ACC4    & ACC5    & Avg  &       & $ACC$ $\rightarrow$ & ACC1    & ACC2    & ACC3    & ACC4    & ACC5    & Avg \\
$\downarrow$ $\hat{\alpha}_{3f}$&     &     &     &     &     &   &       & $\downarrow$ $\hat{\alpha}_{5f}$  &     &     &     &     &     &  \\
			\\
			\hline
			\\
			Q1    & -4.11 & -4.50 & -6.61 & -4.11 & -4.83 & -4.83 &       &    Q1   & -4.25 & -4.21 & -6.42 & -4.25 & -4.67 & -4.76 \\
			& (1.66) & (1.44) & (2.40) & (1.66) & (2.69) & (1.97) &       &       & (1.75) & (1.35) & (2.37) & (1.75) & (2.61) & (1.96) \\
            &       &       &       &       &       &       &       &       &       &       &       &  \\
		Q2	& -1.37 & -1.90 & -1.65 & -1.37 & -3.87 & -2.03 &       &     Q2  & -1.44 & -1.93 & -1.43 & -1.44 & -3.90 & -2.03 \\
			& (0.79) & (0.95) & (0.77) & (0.79) & (2.04) & (1.07) &       &       & (0.85) & (0.99) & (0.68) & (0.85) & (2.05) & (1.08) \\
            &       &       &       &       &       &       &       &       &       &       &       &  \\
		Q3	& -0.99 & -0.20 & 1.19  & -0.99 & -3.58 & -0.91 &       &     Q3  & -1.07 & -0.41 & 0.92  & -1.07 & -3.72 & -1.07 \\
			& (0.54) & (0.14) & (0.70)  & (0.54) & (1.88) & (0.48) &       &       & (0.59) & (0.31) & (0.56)  & (0.59) & (1.95) & (0.58) \\
            &       &       &       &       &       &       &       &       &       &       &       &  \\
		Q4	& -0.33 & 0.05  & 0.99  & -0.33 & -2.29 & -0.38 &       &     Q4  & -0.20 & -0.10 & 0.75  & -0.20 & -2.40 & -0.43 \\
			& (0.18) & (0.04)  & (0.55)  & (0.18) & (1.21) & (0.20) &       &       & (0.11) & (0.08) & (0.43)  & (0.11) & (1.27) & (0.23) \\
            &       &       &       &       &       &       &       &       &       &       &       &  \\
		Q5	& 3.03  & 5.22  & 4.30  & 3.03  & -1.20 & 2.88  &       &     Q5  & 3.50  & 5.25  & 4.16  & 3.50  & -0.92 & 3.10 \\
			& (1.22)  & (1.99)  & (1.65)  & (1.22)  & (0.58) & (1.10)  &       &       & (1.44)  & (2.05)  & (1.64)  & (1.44)  & (0.45) & (1.22) \\
            &       &       &       &       &       &       &       &       &       &       &       &  \\
		\textit{top-bottom}	& 7.14 & 9.73 & 10.91 & 7.14 & 
        3.63 & 7.71 &       &    \textit{top-bottom}   & 7.75 & 9.46 & 10.57 & 7.75 & 3.75 & 7.85 \\
			& (2.60)  & (2.49)  & (3.51)  & (2.60)  & (2.73)  & (2.79)  &       &       & (2.82)  & (2.42)  & (3.39)  & (2.82)  & (2.81)  & (2.85) \\
						&       &       &       &       &       &       &       &       &       &       &       &  \\
			\hline
			\hline
	\end{tabular}}%
	\label{tab:reverse_all}%
\end{table}

\clearpage
\section{Discussion}
\label{discussion}

For comparative purposes we also run the double sorts exercise using for the second sort the $\hat{\delta}^{*}$ measure of managerial skill proposed by \cite{cohen2005judging}. In this specification we obtain prior to the crisis lower average differences than the ones determined by the $ACC$ index (see Table \ref{tab:doublesorting_cohen}), with also more volatile \textit{top-bottom} performances acr$ACC$oss quintiles. For the interval 2007-2010 results of the five-factors model are almost in line with those for the $ACC$ sorting, while in the three-factors model sorting according to $\hat{\delta}^{*}$ produces better extra-performances than those for the $ACC$ index although for both criteria t-statistics do not support significant findings (cf. Table \ref{tab:doublesorting_post}). The overall case for the whole period 2004-2010 is presented in Panel C of Table \ref{tab:doublesorting_cohen}. On average this double-sort criterion determines lower extra-performances than those for the $ACC$ sort, with poor performances for low past-$\hat{\alpha}$ quintiles and less clear monotonic patterns as function of $\hat{\delta}^{*}$ levels.

Similarly to subsection \ref{rs}, we also report in Table \ref{tab:reverse_cohen} the reverse sort quintiles based on $\hat{\delta}^{*}$, getting on average lower annualized extra-performances in each time window. The $\hat{\delta}^{*}$ indicator of \cite{cohen2005judging} can thus be interpreted as an alternative measure of the skills of the manager, while the $ACC$ property can instead be used to refine the ability of the manager in a complementary way to past performances.

\begin{table}[h!]
  \centering
  \caption{\textbf{Double Sorts Funds by Past Performances and $\hat{\delta}^{*}$.} The table shows the resulting 5x5 portfolios' alphas obtained by sorting funds in quintiles firstly by past-$\hat{\alpha}$ performances and then by the $\hat{\delta}^{*}$ measure of managerial skills proposed by \cite{cohen2005judging} for the second sort. The table reports the OLS estimates of each 5x5 portfolio's alpha (in percentage per year) and the corresponding absolute value of the t-statistics (in parentheses). Subscripts $3f$ and $5f$ stand for the three and five factors models used to compute alpha performances (\cite{fama1993common,fama2015five}). To compute past performance we use nine months of lookback period of daily observations on gross returns, which are determined by adding fund management fees to net returns. We then calculate the return of each quintile portfolio over the next three months using daily returns series and equally weighting funds in each 5x5 portfolio. 5x5 portfolios are redefined each quarter and the corresponding three months returns time series are connected across quarters to form a full sample period for each 5x5 portfolio. \textit{top-bottom} is the portfolio obtained investing long in funds belonging to the best-performer quintile portfolio and short funds in the worst-performer quintile, within the same quintile determined in the first sort. Finally, the portfolio denoted as \textit{Avg} invests equally in each of the five \textit{top-bottom} portfolios. Panel A stands for the interval from June 2004 to June 2007, Panel B refers to the period from September 2007 to June 2010, and Panel C from June 2004 to June 2010.}
\scalebox{0.60}{  \begin{tabular}{lccccccp{0.2cm}lcccccc}
\hline
\hline
&       &   & \multicolumn{11}{c}{Panel A (2004-2007): Sort funds by past-$\hat{\alpha}$ and then by $\hat{\delta}^{*}$ of \cite{cohen2005judging}} \\              

$\hat{\alpha}_{3f}$ $\rightarrow$ & Q1    & Q2    & Q3    & Q4    & Q5    & Avg  &       & $\hat{\alpha}_{5f}$ $\rightarrow$ & Q1    & Q2    & Q3    & Q4    & Q5    & Avg \\
$\downarrow$ $\hat{\delta}^{*}$&     &     &     &     &     &   &       & $\downarrow$ $\hat{\delta}^{*}$  &     &     &     &     &     &  \\
    \hline
    \\
    $\hat{\delta}^{*}$1    &     5.77  & 1.62  & 1.91  & 2.57  & 7.67  & 3.91  &       &    $\hat{\delta}^{*}$1     & 2.79  & -0.26 & -0.92 & 0.22  & 6.64  & 1.70 \\
 &   (1.78)  & (1.63)  & (2.19)  & (2.49)  & (2.77)  & (2.17)  &       &       & (0.85)  & (0.25) & (0.95) & (0.23)  & (2.40)  & (0.46) \\
            &       &       &       &       &       &       &       &       &       &       &       &  \\
 
$\hat{\delta}^{*}$2    &     3.90  & 0.03  & 1.39  & 0.88  & 8.73  & 2.98  &       &      $\hat{\delta}^{*}$2   & 1.52  & 0.78  & 0.35  & 2.36  & 8.39  & 2.68 \\
&    (2.05  &  (0.03  & (1.56  & (0.91  & (3.07  & (1.53  &       &       & (0.97)  & (0.85)  & (0.39)  & (2.33)  & (2.65)  & (1.44) \\
           &       &       &       &       &       &       &       &       &       &       &       &  \\
 
$\hat{\delta}^{*}$3    &     3.45  & 0.77  & 1.54  & 1.56  & 7.92  & 3.05  &       &     $\hat{\delta}^{*}$3   & 1.43  & 0.18  & 0.94  & 1.76  & 8.77  & 2.61 \\
&    (1.97)  & (0.73)  & (1.66)  & (1.46)  & (2.47)  & (1.66)  &       &       & (1.10)  & (0.23)  & (1.13)  & (1.75)  & (2.80)  & (1.40) \\
           &       &       &       &       &       &       &       &       &       &       &       &  \\
 
$\hat{\delta}^{*}$4    &     2.92  & 0.97  & 1.89  & 2.04  & 10.14 & 3.59  &       &    $\hat{\delta}^{*}$4     & 0.29  & 1.15  & 1.30  & 0.80  & 11.26 & 2.96 \\
&    (1.79)  & (0.94)  & (1.86)  & (2.02)  & (2.89)  & (1.90)  &       &       & (0.23)  & (1.32)  & (1.77)  & (0.82)  & (3.15)  & (1.46) \\
           &       &       &       &       &       &       &       &       &       &       &       &  \\
 
$\hat{\delta}^{*}$5    &     2.27  & 1.57  & 1.18  & 2.16  & 14.31 & 4.30  &       &   $\hat{\delta}^{*}$5      & 3.24  & 2.37  & 2.88  & 2.26  & 16.58 & 5.47 \\
 &   (1.65)  & (1.62)  & (1.36)  & (1.90)  & (2.60)  & (1.83)  &       &       & (2.31)  & (2.28)  & (2.61)  & (1.55)  & (2.89)  & (2.33) \\
            &       &       &       &       &       &       &       &       &       &       &       &  \\
 
\textit{top-bottom}   &     -3.50  & -0.04  & -0.72  & -0.41  & 6.63 & 0.38 &       &   \textit{top-bottom}    & 0.45 & 2.63 & 3.80 & 2.04 & 9.94 & 3.77 \\
&    (1.35) & (0.07) & (0.99) & (0.57) & (1.83)  & (0.44) &       &       & (0.17)  & (2.27)  & (3.04)  & (1.53)  & (2.43)  & (1.33) \\
             &       &       &       &       &       &       &       &       &       &       &       &  \\
             \hline
            
             &       &   & \multicolumn{11}{c}{Panel B (2007-2010): Sort funds by past-$\hat{\alpha}$ and then by $\hat{\delta}^{*}$ of \cite{cohen2005judging}} \\              
$\hat{\alpha}_{3f}$ $\rightarrow$ & Q1    & Q2    & Q3    & Q4    & Q5    & Avg  &       & $\hat{\alpha}_{5f}$ $\rightarrow$ & Q1    & Q2    & Q3    & Q4    & Q5    & Avg \\
$\downarrow$ $\hat{\delta}^{*}$&     &     &     &     &     &   &       & $\downarrow$ $\hat{\delta}^{*}$  &     &     &     &     &     &  \\
    \hline
    \\
    $\hat{\delta}^{*}$1    & -15.57 & -9.37 & -7.94 & -8.01 & -13.11 & -10.80 &       &     $\hat{\delta}^{*}$1     & -13.71 & -7.63 & -7.27 & -7.40 & -4.45 & -8.09 \\
          & (1.80) & (1.59) & (1.90) & (1.79) & (2.02) & (1.82) &       &       & (1.55) & (1.16) & (1.70) & (1.68) & (1.12) & (1.44) \\
                      &       &       &       &       &       &       &       &       &       &       &       &  \\

   $\hat{\delta}^{*}$2       & -11.29 & -6.47 & -5.92 & -6.38 & -5.03 & -7.02 &       &      $\hat{\delta}^{*}$2   & -9.33 & -5.42 & -4.21 & -5.31 & -3.36 & -5.53 \\
          & (1.43) & (1.56) & (1.69) & (1.45) & (1.07) & (1.44) &       &       & (1.23) & (1.22) & (1.19) & (1.29) & (0.93) & (1.17) \\
                      &       &       &       &       &       &       &       &       &       &       &       &  \\

  $\hat{\delta}^{*}$3        & -8.54 & -8.26 & -7.33 & -4.70 & -3.71 & -6.51 &       &     $\hat{\delta}^{*}$3    & -9.06 & -8.28 & -6.27 & -4.81 & -2.45 & -6.18 \\
          & (1.19) & (2.00) & (2.04) & (1.20) & (0.82) & (1.45) &       &       & (1.33) & (2.12) & (1.84) & (1.25) & (0.61) & (1.43) \\
                      &       &       &       &       &       &       &       &       &       &       &       &  \\

  $\hat{\delta}^{*}$4        & -6.04 & -6.26 & -3.71 & -3.26 & -3.15 & -4.48 &       &      $\hat{\delta}^{*}$4   & -7.48 & -8.21 & -4.25 & -3.62 & -4.44 & -5.60 \\
          & (1.02 & (1.65 & (1.19 & (0.96 & (0.59 & (1.08 &       &       & (1.30) & (2.11) & (1.07) & (1.11) & (0.95) & (1.31) \\
                      &       &       &       &       &       &       &       &       &       &       &       &  \\

 $\hat{\delta}^{*}$5         & -10.74 & -4.74 & -5.84 & -4.80 & -2.24 & -5.67 &       &       $\hat{\delta}^{*}$5  & -11.72 & -4.80 & -5.78 & -3.89 & -3.57 & -5.95 \\
          & (2.20) & (1.21) & (1.74) & (1.29) & (0.31) & (1.35) &       &       & (2.84) & (1.33) & (1.73) & (1.14) & (0.51) & (1.51) \\
                      &       &       &       &       &       &       &       &       &       &       &       &  \\

 \textit{top-bottom}         & 4.83 & 4.62 & 2.11 & 3.21 & 10.88 & 5.13 &       &   \textit{top-bottom}    & 1.98 & 2.82 & 1.50 & 3.51 & 0.88 & 2.14 \\
          & (0.66)  & (0.93)  & (0.61)  & (0.87)  & (1.52)  & (1.26) &       &       & (0.28)  & (0.51)  & (0.42)  & (0.89)  & (0.15)  & (0.55) \\
            &       &       &       &       &       &       &       &       &       &       &       &  \\
            \hline
            &       &   & \multicolumn{11}{c}{Panel C (2004-2010): Sort funds by past-$\hat{\alpha}$ and then by $\hat{\delta}^{*}$ of \cite{cohen2005judging}} \\              
$\hat{\alpha}_{3f}$ $\rightarrow$ & Q1    & Q2    & Q3    & Q4    & Q5    & Avg  &       & $\hat{\alpha}_{5f}$ $\rightarrow$ & Q1    & Q2    & Q3    & Q4    & Q5    & Avg \\
$\downarrow$ $\hat{\delta}^{*}$&     &     &     &     &     &   &       & $\downarrow$ $\hat{\delta}^{*}$  &     &     &     &     &     &  \\
 \hline
 \\
    $\hat{\delta}^{*}$1    & -3.87 & -2.34 & -1.88 & -2.25 & -3.47 & -2.76 &       &   $\hat{\delta}^{*}$1    & -3.32 & -1.88 & -0.98 & -1.99 & 0.24  & -1.58 \\
          & (0.89) & (0.82) & (0.91) & (1.04) & (1.07) & (0.95) &       &       & (0.74) & (0.60) & (0.46) & (0.94) & (0.11)  & (0.53) \\
                       &       &       &       &       &       &       &       &       &       &       &       &  \\

  $\hat{\delta}^{*}$2        & -3.14 & -2.38 & -1.41 & -1.28 & 0.79  & -1.48 &       &   $\hat{\delta}^{*}$2    & -2.89 & -1.72 & -0.82 & -0.91 & 1.98  & -0.87 \\
          & (0.82) & (1.17) & (0.83) & (0.61) & (0.32)  & (0.62) &       &       & (0.79) & (0.80) & (0.47) & (0.47) & (0.93)  & (0.32) \\
                       &       &       &       &       &       &       &       &       &       &       &       &  \\

    $\hat{\delta}^{*}$3      & -2.91 & -3.17 & -2.88 & -1.23 & 2.59  & -1.52 &       &     $\hat{\delta}^{*}$3  & -3.62 & -3.56 & -1.81 & -1.46 & 3.26  & -1.44 \\
          & (0.83) & (1.61) & (1.68) & (0.66) & (0.97)  & (0.76) &       &       & (1.11) & (1.91) & (1.11) & (0.79) & (1.32)  & (0.72) \\
                       &       &       &       &       &       &       &       &       &       &       &       &  \\

    $\hat{\delta}^{*}$4       & -2.36 & -2.61 & -0.58 & 0.06  & 3.62  & -0.37 &       &    $\hat{\delta}^{*}$4   & -3.04 & -3.78 & -1.31 & -0.36 & 3.75  & -0.95 \\
          & (0.83) & (1.42) & (0.39) & (0.04)  & (1.16)  & (0.29) &       &       & (1.12) & (2.05) & (0.70) & (0.22) & (1.28)  & (0.56) \\
                       &       &       &       &       &       &       &       &       &       &       &       &  \\

    $\hat{\delta}^{*}$5       & -4.11 & -0.23 & 0.08  & 1.30  & 7.48  & 0.90  &       &   $\hat{\delta}^{*}$5    & -5.80 & -1.12 & -0.66 & 1.16  & 7.50  & 0.22 \\
          & (1.61) & (0.12) & (0.04)  & (0.63)  & (1.61)  & (0.11)  &       &       & (2.90) & (0.62) & (0.38) & (0.61)  & (1.65)  & (0.33) \\
                       &       &       &       &       &       &       &       &       &       &       &       &  \\

     \textit{top-bottom}      & -0.24  & 2.10 & 1.95 & 3.55 & 10.95 & 3.66 &       &   \textit{top-bottom}    & -2.48  & 0.77 & 0.32 & 3.15 & 7.26 & 1.80 \\
          & (0.06) & (0.85)  & (1.03)  & (1.65)  & (2.58)  & (1.69) &       &       & (0.66) & (0.28)  & (0.16)  & (1.42)  & (1.91)  & (0.86) \\
             &       &       &       &       &       &       &       &       &       &       &       &  \\
          \hline
          \hline
    \end{tabular}}%
  \label{tab:doublesorting_cohen}%
\end{table}

\begin{table}[ht!]
  \centering
  \caption{\textbf{Reverse Double Sorts Funds by  $\hat{\delta}^{*}$ and Past Performances.} The table shows the resulting 5x5 portfolios' alphas obtained by sorting funds in quintiles firstly by $\hat{\delta}^{*}$ and then by past-$\hat{\alpha}$ performances. The table reports the OLS estimates of each 5x5 portfolio's alpha (in percentage per year) and the corresponding absolute value of the t-statistics (in parentheses). Subscripts $3f$ and $5f$ stand for the three and five factors models used to compute alpha performances (\cite{fama1993common,fama2015five}). To compute past performance we use nine months of lookback period of daily observations on gross returns, which are determined by adding fund management fees to net returns. We then calculate the return of each quintile portfolio over the next three months using daily returns series and equally weighting funds in each 5x5 portfolio. 5x5 portfolios are redefined each quarter and the corresponding three months returns time series are connected across quarters to form a full sample period for each 5x5 portfolio. \textit{top-bottom} is the portfolio obtained investing long in funds belonging to the best-performer quintile portfolio and short funds in the worst-performer quintile, within the same quintile determined in the first sort. Finally, the portfolio denoted as \textit{Avg} invests equally in each of the five \textit{top-bottom} portfolios. Panel A stands for the interval from June 2004 to June 2007, Panel B refers to the period from September 2007 to June 2010, and Panel C from June 2004 to June 2010.}
\scalebox{0.57}{  \begin{tabular}{lccccccp{0.2cm}lcccccc}
\hline
\hline
&       &   & \multicolumn{11}{c}{Panel A (Pre Crisis): Sorting funds by $\hat{\delta}^{*}$ of \cite{cohen2005judging} and then by past-$\hat{\alpha}$} \\ 
&     &     &     &     &     &   &       &  &     &     &     &    &   &  \\
$\hat{\delta}^{*}$ $\rightarrow$ & $\hat{\delta}^{*}$1    & $\hat{\delta}^{*}$2    & $\hat{\delta}^{*}$3    & $\hat{\delta}^{*}$4    & $\hat{\delta}^{*}$5    & Avg  &       & $\hat{\delta}^{*}$ $\rightarrow$ & $\hat{\delta}^{*}$1    & $\hat{\delta}^{*}$2    & $\hat{\delta}^{*}$3    & $\hat{\delta}^{*}$4    & $\hat{\delta}^{*}$5    & Avg \\
$\downarrow$ $\hat{\alpha}_{3f}$&     &     &     &     &     &   &       & $\downarrow$ $\hat{\alpha}_{5f}$  &     &     &     &     &     &  \\
 \hline
 \\
    Q1    & 8.63  & 1.58  & 1.56  & 2.79  & 10.67 & 5.04  &       &   Q1    & 4.08  & 0.57  & -0.90 & -0.26 & 9.02  & 2.50 \\
          & (1.85)  & (1.04)  & (1.38)  & (2.88)  & (2.91)  & (2.01)  &       &       & (0.89)  & (0.34)  & (0.69) & (0.26) & (2.39)  & (0.53) \\
                       &       &       &       &       &       &       &       &       &       &       &       &  \\

Q2          & 2.73  & 1.01  & 0.77  & -0.93 & 10.18 & 2.75  &       &   Q2    & 0.34  & 0.60  & 1.36  & 0.25  & 11.81 & 2.87 \\
          & (1.03)  & (0.65)  & (0.72)  & (0.90) & (2.73)  & (0.84)  &       &       & (0.14)  & (0.51)  & (1.39)  & (0.24)  & (3.03)  & (1.06) \\
                       &       &       &       &       &       &       &       &       &       &       &       &  \\

Q3          & 5.80  & 2.74  & 1.25  & 1.61  & 12.17 & 4.71  &       &   Q3    & 1.99  & -1.24 & -0.35 & 0.42  & 12.84 & 2.73 \\
          & (2.96)  & (1.86)  & (0.96)  & (1.74)  & (2.98)  & (2.10)  &       &       & (1.20)  & (1.03) & (0.37) & (0.52)  & (3.04)  & (0.67)) \\
                       &       &       &       &       &       &       &       &       &       &       &       &  \\

 Q4         & 3.51  & 2.19  & 1.33  & 0.80  & 11.24 & 3.81  &       &   Q4    & 1.79  & 1.39  & 1.49  & 1.39  & 11.42 & 3.50 \\
          & (1.81)  & (1.35)  & (1.13)  & (0.88)  & (2.57)  & (1.55)  &       &       & (1.08)  & (1.14)  & (1.61)  & (1.43)  & (2.48)  & (1.55) \\
                       &       &       &       &       &       &       &       &       &       &       &       &  \\

 Q5         & 2.81  & 2.09  & 1.70  & 1.20  & 16.29 & 4.82  &       &    Q5   & 4.41  & 2.51  & 2.41  & 2.64  & 19.16 & 6.23 \\
          & (1.56)  & (1.40)  & (1.44)  & (1.26)  & (2.55)  & (1.64)  &       &       & (2.56)  & (1.56)  & (1.92)  & (2.05)  & (2.89)  & (2.20) \\
                       &       &       &       &       &       &       &       &       &       &       &       &  \\

 top-bottom         & -5.82  & 0.51 & 0.14 & -1.59  & 5.63 & -0.23  &       &    top-bottom   & 0.34 & 1.93 & 3.32 & 2.90 & 10.14 & 3.72 \\
          & (1.46) & (0.43)  & (0.15)  & (1.67) & (1.38)  & (0.18)  &       &       & (0.08)  & (1.03)  & (2.15)  & (2.05)  & (2.15)  & (2.58) \\
             &       &       &       &       &       &       &       &       &       &       &       &  \\
   \hline          
             &       &   & \multicolumn{11}{c}{Panel B (Crisis): Sorting funds by $\hat{\delta}^{*}$ of \cite{cohen2005judging} and then by past-$\hat{\alpha}$} \\ 
&     &     &     &     &     &   &       &  &     &     &     &    &   &  \\
$\hat{\delta}^{*}$ $\rightarrow$ & $\hat{\delta}^{*}$1    & $\hat{\delta}^{*}$2    & $\hat{\delta}^{*}$3    & $\hat{\delta}^{*}$4    & $\hat{\delta}^{*}$5    & Avg  &       & $\hat{\delta}^{*}$ $\rightarrow$ & $\hat{\delta}^{*}$1    & $\hat{\delta}^{*}$2    & $\hat{\delta}^{*}$3    & $\hat{\delta}^{*}$4    & $\hat{\delta}^{*}$5    & Avg \\
$\downarrow$ $\hat{\alpha}_{3f}$&     &     &     &     &     &   &       & $\downarrow$ $\hat{\alpha}_{5f}$  &     &     &     &     &     &  \\
 \hline
 \\
    Q1    & -13.17 & -7.28 & -4.85 & -7.08 & -5.59 & -7.59 &       &    Q1   & -20.71 & -9.09 & -10.29 & -7.36 & -7.27 & -10.94 \\
          & (1.56) & (1.86) & (1.43) & (1.82) & (1.19) & (1.57) &       &       & (2.29) & (1.81) & (2.56) & (1.89) & (1.53) & (2.02) \\
                      &       &       &       &       &       &       &       &       &       &       &       &  \\

Q2          & -9.79 & -6.88 & -6.99 & -2.97 & -5.28 & -6.38 &       &   Q2    & -8.11 & -7.58 & -7.55 & -2.90 & -1.78 & -5.58 \\
          & (1.33) & (1.78) & (1.77) & (0.90) & (1.13) & (1.38) &       &       & (1.22) & (1.83) & (1.99) & (0.84) & (0.43) & (1.26) \\
                      &       &       &       &       &       &       &       &       &       &       &       &  \\

 Q3         & -9.07 & -6.20 & -3.43 & -5.80 & -4.72 & -5.84 &       &   Q3    & -6.50 & -6.56 & -5.24 & -3.46 & -3.44 & -5.04 \\
          & (1.37) & (1.74) & (0.96) & (1.61) & (0.99) & (1.33) &       &       & (0.97) & (1.90) & (1.42) & (1.01) & (0.96) & (1.25) \\
                      &       &       &       &       &       &       &       &       &       &       &       &  \\

   Q4       & -8.01 & -4.62 & -6.40 & -4.24 & -4.86 & -5.62 &       &    Q4   & -8.31 & -4.82 & -5.30 & -3.79 & -1.50 & -4.74 \\
          & (1.40) & (1.23) & (1.80) & (1.19) & (0.86) & (1.30) &       &       & (1.26) & (1.37) & (1.27) & (1.05) & (0.34) & (1.06) \\
                      &       &       &       &       &       &       &       &       &       &       &       &  \\

     Q5     & -11.05 & -5.67 & -5.98 & -5.97 & -3.79 & -6.49 &       &     Q5  & -6.27 & -6.29 & -5.91 & -3.94 & -4.29 & -5.34 \\
          & (1.84) & (1.51) & (1.54) & (1.68) & (0.50) & (1.41) &       &       & (1.16) & (1.48) & (1.60) & (1.09) & (0.61) & (1.19) \\
                      &       &       &       &       &       &       &       &       &       &       &       &  \\

    top-bottom      & 2.12 & 1.61 & -1.13  & 1.11 & 1.80 & 1.10 &       &    top-bottom   & 14.43 & 2.81 & 4.38 & 3.42 & 2.98 & 5.60 \\
          & (0.39)  & (1.35)  & (1.02) & (0.90)  & (0.44)  & (0.74) &       &       & (2.41)  & (0.96)  & (2.06)  & (1.34)  & (0.57)  & (2.26) \\
             &       &       &       &       &       &       &       &       &       &       &       &  \\
             \hline
             &       &   & \multicolumn{11}{c}{Panel B: Sorting funds by $\hat{\delta}^{*}$ of \cite{cohen2005judging} and then by past-$\hat{\alpha}$} \\ 
&     &     &     &     &     &   &       &  &     &     &     &    &   &  \\
$\hat{\delta}^{*}$ $\rightarrow$ & $\hat{\delta}^{*}$1    & $\hat{\delta}^{*}$2    & $\hat{\delta}^{*}$3    & $\hat{\delta}^{*}$4    & $\hat{\delta}^{*}$5    & Avg  &       & $\hat{\delta}^{*}$ $\rightarrow$ & $\hat{\delta}^{*}$1    & $\hat{\delta}^{*}$2    & $\hat{\delta}^{*}$3    & $\hat{\delta}^{*}$4    & $\hat{\delta}^{*}$5    & Avg \\
$\downarrow$ $\hat{\alpha}_{3f}$&     &     &     &     &     &   &       & $\downarrow$ $\hat{\alpha}_{5f}$  &     &     &     &     &     &  \\
 \hline
 \\
    Q1    & -2.46 & -2.60 & -1.02 & -1.68 & 1.93  & -1.17 &       &    Q1   & -7.51 & -4.38 & -5.06 & -2.96 & 0.00  & -3.98 \\
          & (0.57) & (1.37) & (0.62) & (0.89) & (0.72)  & (0.55) &       &       & (1.66) & (1.84) & (2.64) & (1.57) & (0.00)  & (1.54) \\
                       &       &       &       &       &       &       &       &       &       &       &       &  \\

  Q2          & -2.05 & -3.13 & -2.22 & -0.60 & 2.49  & -1.10 &       &   Q2    & -2.31 & -3.05 & -3.05 & 0.16  & 3.83  & -0.89 \\
          & (0.57) & (1.67) & (1.16) & (0.38) & (0.93)  & (0.57) &       &       & (0.72) & (1.55) & (1.68) & (0.09)  & (1.46)  & (0.48) \\
                       &       &       &       &       &       &       &       &       &       &       &       &  \\

  Q3          & -1.71 & -2.38 & -0.56 & -1.38 & 2.23  & -0.76 &       &     Q3  & -1.57 & -2.78 & -1.53 & -0.31 & 3.14  & -0.61 \\
          & (0.53) & (1.35) & (0.32) & (0.78) & (0.79)  & (0.44) &       &       & (0.49) & (1.67) & (0.87) & (0.19) & (1.29)  & (0.38) \\
                       &       &       &       &       &       &       &       &       &       &       &       &  \\

  Q4          & -1.81 & -1.57 & -1.62 & -0.50 & 3.20  & -0.46 &       &     Q4  & -3.22 & -1.40 & -1.45 & -0.92 & 5.02  & -0.39 \\
          & (0.64) & (0.85) & (0.93) & (0.29) & (0.98)  & (0.35) &       &       & (1.03) & (0.82) & (0.74) & (0.53) & (1.75)  & (0.27) \\
                       &       &       &       &       &       &       &       &       &       &       &       &  \\

  Q5          & -3.62 & -1.88 & -1.93 & -1.28 & 5.84  & -0.57 &       &     Q5  & -0.37 & -1.21 & -0.64 & -0.01 & 7.61  & 1.08 \\
          & (1.26) & (1.03) & (1.03) & (0.73) & (1.27)  & (0.56) &       &       & (0.14) & (0.59) & (0.35) & (0.01) & (1.69)  & (0.12) \\
                       &       &       &       &       &       &       &       &       &       &       &       &  \\

  top-bottom        & -1.16  & 0.72 & -0.91  & 0.40 & 3.92 & 0.59 &       &  top-bottom     & 7.14 & 3.17 & 4.43 & 2.95 & 7.61 & 5.06 \\
          & (0.41) & (1.11)  & (1.41) & (0.58)  & (1.42)  & (0.70) &       &       & (2.28)  & (2.12)  & (3.64)  & (2.14)  & (2.27)  & (2.41) \\
             &       &       &       &       &       &       &       &       &       &       &       &  \\

          \hline
          \hline
    \end{tabular}}%
  \label{tab:reverse_cohen}%
\end{table}

From an investor perspective it would be valuable to combine therefore the information present in both $ACC$ and past-$\hat{\alpha}$ to build portfolios. The highest performances in our cases are offered for instance by Q5ACC1-Q1ACC5 prior to the crisis (Panel A of Table \ref{tab:doublesorting_pre}) which would get annualized extra-performances equal to about [17.94; 19.70] per cent. A similar strategy would be profitable also when the reference period is the entire interval from 2004 to 2010; interestingly, even circumscribing to the crisis period 2007-2010 only, following this investment style would have helped to mitigate those poor performances instead observed in most of the quintile portfolios. These alphas are, therefore, usually higher than those obtained using only a single quintile sort, suggesting that investors would benefit from combining both sources of information in constructing portfolios. 

Similar practical results can be drawn for the reverse sorts of Tables \ref{tab:reverse_pre}-\ref{tab:reverse_all}. For instance, for the entire period 2004-2010 the strategy ACC1Q5-ACC5Q1 would generate annual performances of about [12.70; 11.65] per cent, which represent again values higher than the average one-way quintile sort.

The $ACC$ index of a portfolio does not emerge simply as a proxy for managerial skills, but rather as a topological alternative to the diversification dimension of an investment strategy. For instance, investors can benefit from this topological information by combining the desired level of the $ACC$ index and selecting skilled managers according to past-$\hat{\alpha}$. Our analysis provides some guidelines for this decision: over a period affected by a boom and bust cycle, managers got better extra-performances by investing in less common assets, and those managers more skilled (namely, with higher past-$\hat{\alpha}$) seem to be the ones that mostly gained from extracting information from this topological feature. 

\clearpage
\section{Conclusion}
\label{conclusion}

This paper advances a bipartite network representation of the funds-constituents relationships to extract valuable information from mutual funds' portfolio compositions. The topological investigation of the system via network centrality measures helps in identifying not only those funds that diversify the most in terms of portfolio composition, but it is also useful in recognizing either those assets that are present in a huge share of funds or, alternatively, those that are held by few portfolios only. Thus, for a given level of diversification, these measures discriminate between those funds more prone to invest in niche markets and those that opt for common assets.

Our findings point to a negative relationship between funds' extra-performances and the average popularity of the assets held in the portfolios, meaning that those funds investing in less popular assets were more likely to produce positive extra-performances in the period 2003-2010. These more niche investment positions might have been less impacted by fire sales arising due to the financial turmoil that spread after mid-2007, thus limiting negative triggering effects in the markets. The topological information gained from portfolio holdings thus emerges as a complementary source of information that can be combined with past alpha measures to better discriminate among funds. 

We propose to exploit the information behind these cross-holdings to built profitable investment strategies that combine both past alpha information, as a signal for persistence in managerial skills, and the topological features of the assets, which mimic actual diversification through more or less popular/common stocks in the market.

From an investor perspective the $ACC$ index can be interpreted as an alternative measure for diversification which takes into account the popularity of the assets across funds' portfolios, offering therefore a competitive view on the actual extent of diversification related to certain portfolio holdings. 

\section*{Acknowledgments}
Authors acknowledge support from CNR PNR Project ``CRISIS Lab''.
\newpage
\bibliographystyle{apalike}
\bibliography{biblio.bib}

\begin{thebibliography}{42}
\providecommand{\natexlab}[1]{#1}
\providecommand{\url}[1]{\texttt{#1}}
\expandafter\ifx\csname urlstyle\endcsname\relax
  \providecommand{\doi}[1]{doi: #1}\else
  \providecommand{\doi}{doi: \begingroup \urlstyle{rm}\Url}\fi

\bibitem[Allen et~al.(2012)Allen, Babus, and Carletti]{allen2012asset}
Franklin Allen, Ana Babus, and Elena Carletti.
\newblock Asset commonality, debt maturity and systemic risk.
\newblock \emph{Journal of Financial Economics}, 104\penalty0 (3):\penalty0
  519--534, 2012.

\bibitem[Alshamsi et~al.(2018)Alshamsi, Pinheiro, and
  Hidalgo]{alshamsi2018optimal}
Aamena Alshamsi, Fl{\'a}vio~L Pinheiro, and Cesar~A Hidalgo.
\newblock Optimal diversification strategies in the networks of related
  products and of related research areas.
\newblock \emph{Nature communications}, 9\penalty0 (1):\penalty0 1328, 2018.

\bibitem[Bahar et~al.(2014)Bahar, Hausmann, and Hidalgo]{bahar2014neighbors}
Dany Bahar, Ricardo Hausmann, and Cesar~A Hidalgo.
\newblock Neighbors and the evolution of the comparative advantage of nations:
  Evidence of international knowledge diffusion?
\newblock \emph{Journal of International Economics}, 92\penalty0 (1):\penalty0
  111--123, 2014.

\bibitem[Balassa(1965)]{balassa1965trade}
Bela Balassa.
\newblock Trade liberalisation and “revealed” comparative advantage.
\newblock \emph{The Manchester School}, 33\penalty0 (2):\penalty0 99--123,
  1965.

\bibitem[Barras et~al.(2010)Barras, Scaillet, and Wermers]{barras2010false}
Laurent Barras, Olivier Scaillet, and Russ Wermers.
\newblock False discoveries in mutual fund performance: Measuring luck in
  estimated alphas.
\newblock \emph{The Journal of Finance}, 65\penalty0 (1):\penalty0 179--216,
  2010.

\bibitem[Barucca and Lillo(2016)]{barucca2016disentangling}
Paolo Barucca and Fabrizio Lillo.
\newblock Disentangling bipartite and core-periphery structure in financial
  networks.
\newblock \emph{Chaos, Solitons \& Fractals}, 88:\penalty0 244--253, 2016.

\bibitem[Bethke et~al.(2017)Bethke, Gehde-Trapp, and Kempf]{bethke2017investor}
Sebastian Bethke, Monika Gehde-Trapp, and Alexander Kempf.
\newblock Investor sentiment, flight-to-quality, and corporate bond comovement.
\newblock \emph{Journal of Banking \& Finance}, 82:\penalty0 112--132, 2017.

\bibitem[Bollen and Busse(2004)]{bollen2004short}
Nicolas~PB Bollen and Jeffrey~A Busse.
\newblock Short-term persistence in mutual fund performance.
\newblock \emph{The Review of Financial Studies}, 18\penalty0 (2):\penalty0
  569--597, 2004.

\bibitem[Brown and Goetzmann(1995)]{brown1995performance}
Stephen~J Brown and William~N Goetzmann.
\newblock Performance persistence.
\newblock \emph{The Journal of Finance}, 50\penalty0 (2):\penalty0 679--698,
  1995.

\bibitem[Busse et~al.(2010)Busse, Goyal, and Wahal]{busse2010performance}
Jeffrey~A Busse, Amit Goyal, and Sunil Wahal.
\newblock Performance and persistence in institutional investment management.
\newblock \emph{The Journal of Finance}, 65\penalty0 (2):\penalty0 765--790,
  2010.

\bibitem[Caccioli et~al.(2014)Caccioli, Shrestha, Moore, and
  Farmer]{caccioli2014stability}
Fabio Caccioli, Munik Shrestha, Cristopher Moore, and J~Doyne Farmer.
\newblock Stability analysis of financial contagion due to overlapping
  portfolios.
\newblock \emph{Journal of Banking \& Finance}, 46:\penalty0 233--245, 2014.

\bibitem[Carhart(1997)]{carhart1997persistence}
Mark~M Carhart.
\newblock On persistence in mutual fund performance.
\newblock \emph{The Journal of Finance}, 52\penalty0 (1):\penalty0 57--82,
  1997.

\bibitem[Cohen et~al.(2005)Cohen, Coval, and P{\'a}stor]{cohen2005judging}
Randolph~B Cohen, Joshua~D Coval, and L'ubo{\v{s}} P{\'a}stor.
\newblock Judging fund managers by the company they keep.
\newblock \emph{The Journal of Finance}, 60\penalty0 (3):\penalty0 1057--1096,
  2005.

\bibitem[Corsi et~al.(2016)Corsi, Marmi, and Lillo]{Corsi}
Fulvio Corsi, Stefano Marmi, and Fabrizio Lillo.
\newblock When micro prudence increases macro risk: The destabilizing effects
  of financial innovation, leverage, and diversification.
\newblock \emph{Operations Research}, 64:\penalty0 1073--1088, 2016.

\bibitem[Coval and Moskowitz(1999)]{coval1999home}
Joshua~D Coval and Tobias~J Moskowitz.
\newblock Home bias at home: Local equity preference in domestic portfolios.
\newblock \emph{The Journal of Finance}, 54\penalty0 (6):\penalty0 2045--2073,
  1999.

\bibitem[Cremers and Petajisto(2009)]{cremers2009active}
KJ~Martijn Cremers and Antti Petajisto.
\newblock How active is your fund manager? a new measure that predicts
  performance.
\newblock \emph{The Review of Financial Studies}, 22\penalty0 (9):\penalty0
  3329--3365, 2009.

\bibitem[Desmarchelier et~al.(2018)Desmarchelier, Regis, and
  Salike]{desmarchelier2018product}
Beno{\^\i}t Desmarchelier, Paulo~Jos{\'e} Regis, and Nimesh Salike.
\newblock Product space and the development of nations: A model of product
  diversification.
\newblock \emph{Journal of Economic Behavior \& Organization}, 145:\penalty0
  34--51, 2018.

\bibitem[Di~Gangi et~al.(2018)Di~Gangi, Lillo, and Pirino]{DiGangi}
Domenico Di~Gangi, Fabrizio Lillo, and Davide Pirino.
\newblock Assessing systemic risk due to fire sales spillover through maximum
  entropy network reconstruction.
\newblock \emph{Journal of Economic Dynamics and Control}, 94:\penalty0
  117--141, 2018.

\bibitem[Elton et~al.(1996)Elton, Gruber, and Blake]{elton1996persistence}
Edwin~J Elton, Martin~J Gruber, and Christopher~R Blake.
\newblock The persistence of risk-adjusted mutual fund performance.
\newblock \emph{Journal of Business}, pages 133--157, 1996.

\bibitem[Fama and French(1993)]{fama1993common}
Eugene~F Fama and Kenneth~R French.
\newblock Common risk factors in the returns on stocks and bonds.
\newblock \emph{Journal of Financial Economics}, 33\penalty0 (1):\penalty0
  3--56, 1993.

\bibitem[Fama and French(2010)]{fama2010luck}
Eugene~F Fama and Kenneth~R French.
\newblock Luck versus skill in the cross-section of mutual fund returns.
\newblock \emph{The Journal of Finance}, 65\penalty0 (5):\penalty0 1915--1947,
  2010.

\bibitem[Fama and French(2015)]{fama2015five}
Eugene~F Fama and Kenneth~R French.
\newblock A five-factor asset pricing model.
\newblock \emph{Journal of Financial Economics}, 116\penalty0 (1):\penalty0
  1--22, 2015.

\bibitem[Flannery and James(1984)]{flannery1984effect}
Mark~J Flannery and Christopher~M James.
\newblock The effect of interest rate changes on the common stock returns of
  financial institutions.
\newblock \emph{The Journal of Finance}, 39\penalty0 (4):\penalty0 1141--1153,
  1984.

\bibitem[Gala et~al.(2017)Gala, Camargo, and Freitas]{gala2017economic}
Paulo Gala, Jhean Camargo, and Elton Freitas.
\newblock The economic commission for latin america and the caribbean (eclac)
  was right: scale-free complex networks and core-periphery patterns in world
  trade.
\newblock \emph{Cambridge Journal of Economics}, 42\penalty0 (3):\penalty0
  633--651, 2017.

\bibitem[Goetzmann and Ibbotson(1994)]{goetzmann1994winners}
William~N Goetzmann and Roger~G Ibbotson.
\newblock Do winners repeat?
\newblock \emph{Journal of Portfolio Management}, 20\penalty0 (2):\penalty0
  9--18, 1994.

\bibitem[Greenwood et~al.(2015)Greenwood, Landier, and Thesmar]{Greenwood}
R.~Greenwood, A.~Landier, and D.~Thesmar.
\newblock Vulnerable banks.
\newblock \emph{Journal of Financial Economics}, 115:\penalty0 471--485, 2015.

\bibitem[Grinblatt and Titman(1992)]{grinblatt1992persistence}
Mark Grinblatt and Sheridan Titman.
\newblock The persistence of mutual fund performance.
\newblock \emph{The Journal of Finance}, 47\penalty0 (5):\penalty0 1977--1984,
  1992.

\bibitem[Hartmann et~al.(2017)Hartmann, Guevara, Jara-Figueroa, Aristar{\'a}n,
  and Hidalgo]{hartmann2017linking}
Dominik Hartmann, Miguel~R Guevara, Cristian Jara-Figueroa, Manuel
  Aristar{\'a}n, and C{\'e}sar~A Hidalgo.
\newblock Linking economic complexity, institutions, and income inequality.
\newblock \emph{World Development}, 93:\penalty0 75--93, 2017.

\bibitem[Hendricks et~al.(1993)Hendricks, Patel, and
  Zeckhauser]{hendricks1993hot}
Darryll Hendricks, Jayendu Patel, and Richard Zeckhauser.
\newblock Hot hands in mutual funds: Short-run persistence of relative
  performance, 1974--1988.
\newblock \emph{The Journal of Finance}, 48\penalty0 (1):\penalty0 93--130,
  1993.

\bibitem[Hidalgo and Hausmann(2009)]{hidalgo2009building}
C{\'e}sar~A Hidalgo and Ricardo Hausmann.
\newblock The building blocks of economic complexity.
\newblock \emph{Proceedings of the national academy of sciences}, 106\penalty0
  (26):\penalty0 10570--10575, 2009.

\bibitem[Hidalgo et~al.(2007)Hidalgo, Klinger, Barab{\'a}si, and
  Hausmann]{hidalgo2007product}
C{\'e}sar~A Hidalgo, Bailey Klinger, A-L Barab{\'a}si, and Ricardo Hausmann.
\newblock The product space conditions the development of nations.
\newblock \emph{Science}, 317\penalty0 (5837):\penalty0 482--487, 2007.

\bibitem[Huang et~al.(2013)Huang, Vodenska, Havlin, and
  Stanley]{huang2013cascading}
Xuqing Huang, Irena Vodenska, Shlomo Havlin, and H~Eugene Stanley.
\newblock Cascading failures in bi-partite graphs: model for systemic risk
  propagation.
\newblock \emph{Scientific reports}, 3:\penalty0 1219, 2013.

\bibitem[Kacperczyk and Schnabl(2010)]{kacperczyk2010safe}
Marcin Kacperczyk and Philipp Schnabl.
\newblock When safe proved risky: Commercial paper during the financial crisis
  of 2007-2009.
\newblock \emph{Journal of Economic Perspectives}, 24\penalty0 (1):\penalty0
  29--50, 2010.

\bibitem[Kacperczyk et~al.(2005)Kacperczyk, Sialm, and
  Zheng]{kacperczyk2005industry}
Marcin Kacperczyk, Clemens Sialm, and Lu~Zheng.
\newblock On the industry concentration of actively managed equity mutual
  funds.
\newblock \emph{The Journal of Finance}, 60\penalty0 (4):\penalty0 1983--2011,
  2005.

\bibitem[Morrison et~al.(2017)Morrison, Buldyrev, Imbruno, Arrieta, Rungi,
  Riccaboni, and Pammolli]{morrison2017economic}
Greg Morrison, Sergey~V Buldyrev, Michele Imbruno, Omar Alonso~Doria Arrieta,
  Armando Rungi, Massimo Riccaboni, and Fabio Pammolli.
\newblock On economic complexity and the fitness of nations.
\newblock \emph{Scientific Reports}, 7\penalty0 (1):\penalty0 15332, 2017.

\bibitem[Namvar and Phillips(2013)]{namvar2013commonalities}
Ethan Namvar and Blake Phillips.
\newblock Commonalities in investment strategy and the determinants of
  performance in mutual fund mergers.
\newblock \emph{Journal of Banking \& Finance}, 37\penalty0 (2):\penalty0
  625--635, 2013.

\bibitem[P{\'a}stor and Stambaugh(2002)]{pastor2002mutual}
L'ubo{\v{s}} P{\'a}stor and Robert~F Stambaugh.
\newblock Mutual fund performance and seemingly unrelated assets.
\newblock \emph{Journal of Financial Economics}, 63\penalty0 (3):\penalty0
  315--349, 2002.

\bibitem[R{\"o}sch and Kaserer(2013)]{rosch2013market}
Christoph~G R{\"o}sch and Christoph Kaserer.
\newblock Market liquidity in the financial crisis: The role of liquidity
  commonality and flight-to-quality.
\newblock \emph{Journal of Banking \& Finance}, 37\penalty0 (7):\penalty0
  2284--2302, 2013.

\bibitem[Schwarzkopf and Farmer(2010)]{schwarzkopf2010}
Yonathan Schwarzkopf and J~Doyne Farmer.
\newblock Empirical study of the tails of mutual fund size.
\newblock \emph{Physical Review E}, 81\penalty0 (6):\penalty0 066113, 2010.

\bibitem[Tacchella et~al.(2012)Tacchella, Cristelli, Caldarelli, Gabrielli, and
  Pietronero]{tacchella2012new}
Andrea Tacchella, Matthieu Cristelli, Guido Caldarelli, Andrea Gabrielli, and
  Luciano Pietronero.
\newblock A new metrics for countries' fitness and products' complexity.
\newblock \emph{Scientific reports}, 2:\penalty0 723, 2012.

\bibitem[Tacchella et~al.(2013)Tacchella, Cristelli, Caldarelli, Gabrielli, and
  Pietronero]{tacchella2013economic}
Andrea Tacchella, Matthieu Cristelli, Guido Caldarelli, Andrea Gabrielli, and
  Luciano Pietronero.
\newblock Economic complexity: conceptual grounding of a new metrics for global
  competitiveness.
\newblock \emph{Journal of Economic Dynamics and Control}, 37\penalty0
  (8):\penalty0 1683--1691, 2013.

\bibitem[Tumminello et~al.(2011)Tumminello, Micciche, Lillo, Piilo, and
  Mantegna]{tumminello2011statistically}
Michele Tumminello, Salvatore Micciche, Fabrizio Lillo, Jyrki Piilo, and
  Rosario~N Mantegna.
\newblock Statistically validated networks in bipartite complex systems.
\newblock \emph{PloS one}, 6\penalty0 (3):\penalty0 e17994, 2011.

\end{thebibliography}

\end{document}